\documentclass[reprent,aps,prb,twocolumn,
superscriptaddress,nofootinbib]{revtex4-2}
\usepackage{amsmath,amssymb}
\usepackage{graphicx}
\usepackage{bm}
\usepackage{multirow}
\usepackage{hyperref}
\usepackage{braket}

\usepackage{color}

\allowdisplaybreaks[2]

\begin{document}

\title{
Anisotropic skyrmion crystal on a centrosymmetric square lattice\\ under an in-plane magnetic field
}
\author{Satoru Hayami}
\email{hayami@phys.sci.hokudai.ac.jp}
\affiliation{Graduate School of Science, Hokkaido University, Sapporo 060-0810, Japan}

\begin{abstract}
We investigate the instability toward a skyrmion crystal under an in-plane magnetic field in centrosymmetric tetragonal magnets. 
By performing the simulated annealing for a spin model on the square lattice, we find that the interplay among the easy-plane two-spin anisotropic interaction, positive biquadratic interaction, and high-harmonic wave-vector interaction plays an important role in stabilizing the skyrmion crystal in the in-plane external magnetic field. 
The obtained skyrmion crystal consists of a pair of vortex and antivortex with opposite polarization in terms of the out-of-plane spin component. 
We also show that various multiple-$Q$ states emerge depending on the model parameters. 
Our results indicate the possibility of topological spin textures in centrosymmetric tetragonal magnets with easy-plane magnetic anisotropy when the ground-state spin configuration at zero field is characterized by a single-$Q$ in-plane cycloidal spiral state. 
\end{abstract}

\maketitle

\section{Introduction}

A magnetic skyrmion, which is characterized by a swirling spin texture so as to have an integer topological (skyrmion) number, has attracted many researchers owing to its rich physical properties~\cite{Bogdanov94, bocdanov1994properties, bogdanov1999stability, nagaosa2013topological, finocchio2016magnetic, fert2017magnetic, zhang2020skyrmion}. 
Through extensive studies in both theory and experiments, a magnetic skyrmion crystal (SkX), which is a periodic array of the magnetic skyrmion, has been found in materials with both noncentrosymmetric and centrosymmetric lattice structures~\cite{Muhlbauer_2009skyrmion,yu2010real,yu2011near,seki2012observation, heinze2011spontaneous, Adams2012, kurumaji2019skyrmion,khanh2020nanometric, Tokura_doi:10.1021/acs.chemrev.0c00297}, and their stabilization mechanisms have been elucidated; the Dzyaloshinskii-Moriya (DM) interaction becomes an important source of the SkX in noncentrosymmetric magnets~\cite{dzyaloshinsky1958thermodynamic,moriya1960anisotropic,rossler2006spontaneous, Yi_PhysRevB.80.054416}, while frustrated exchange interactions, multiple spin interactions, and/or symmetric anisotropic interactions play an important role in inducing the SkX in centrosymmetric magnets~\cite{Okubo_PhysRevLett.108.017206, leonov2015multiply, batista2016frustration, hayami2021topological, Yambe_PhysRevB.106.174437}. 

In addition to the exchange interactions that are inherent to the materials, an external magnetic field is another essential ingredient to induce the SkX, where the instability toward the SkX depends on the magnetic field direction; the magnetic field perpendicular to the ordering wave vectors tends to stabilize the SkX~\cite{rossler2006spontaneous, Yi_PhysRevB.80.054416, hayami2021field}. 
Meanwhile, it was shown that the SkX and similar topological spin crystals can be stabilized even in the magnetic field parallel to the ordering wave vectors. 
For example, a magnetic bimeron crystal, which is formed by a periodic alignment of a pair of two merons, can appear under the in-plane magnetic field in noncentrosymmetric magnets with the DM interaction~\cite{Borge_PhysRevB.99.060407, Moon_PhysRevApplied.12.064054}. 
Another example is the SkX in centrosymmetric magnets, which is stabilized without the DM interaction by considering the easy-plane magnetic anisotropy and a triangular-lattice geometry~\cite{Hayami_PhysRevB.103.224418, Hayami_PhysRevB.103.054422}. 
Meanwhile, it has not been clarified the possibility of the SkXs under the in-plane field on a centrosymmetric square lattice, where the effect of geometrical frustration does not appear. 
Although the emergence of the SkXs on such a square lattice in the out-of-plane magnetic field has been shown by the theoretical model calculations~\cite{Hayami_PhysRevB.103.024439, Utesov_PhysRevB.103.064414, Wang_PhysRevB.103.104408, Hayami_PhysRevB.105.174437} and experiments in GdRu$_2$Si$_2$~\cite{khanh2020nanometric, Yasui2020imaging, khanh2022zoology, Matsuyama_PhysRevB.107.104421, hayami2023widely, Wood_PhysRevB.107.L180402} and EuAl$_4$~\cite{takagi2022square, Zhu_PhysRevB.105.014423, hayami2023orthorhombic, Gen_PhysRevB.107.L020410}, its stabilization condition under the in-plane magnetic field has been unclear yet. 

In the present study, we investigate the stability of the in-plane field-induced SkX in the centrosymmetric tetragonal system. 
Through the numerical calculations based on the simulated annealing for the phenomenological spin model, we find the conditions to stabilize the SkX on the square lattice under the in-plane field. 
We show that the interplay among the easy-plane two-spin anisotropic interaction, positive biquadratic interaction, and high-harmonic wave-vector interaction is important in order to realize the SkX. 
The obtained SkX is characterized by an anisotropic double-$Q$ spin modulation, which reflects the lack of the fourfold rotational symmetry under the in-plane field.  
We also find various double-$Q$ instabilities depending on the magnitude of the magnetic field and easy-axis anisotropic interaction. 

The rest of this paper is organized as follows. 
In Sec.~\ref{sec: Model and method}, we present the effective spin model consisting of bilinear and biquadratic spin interactions on the square lattice. 
We also outline the numerical simulated annealing. 
In Sec.~\ref{sec: Results}, we show the results of the magnetic phase diagram at low temperatures. 
We describe the spin and scalar chirality configurations in real and momentum spaces in each phase. 
Section~\ref{sec: Summary} is devoted to a summary of this paper.  
In Appendix~\ref{sec: Results for other model parameters}, we show two magnetic phase diagrams calculated by using different model parameters in the main text. 

\section{Model and method}
\label{sec: Model and method}

\subsection{Model}

In order to investigate the essence in terms of the stability of the in-plane field-induced SkX at low temperatures, we analyze the spin model with the momentum-resolved interaction on the two-dimensional centrosymmetric square lattice~\cite{hayami2021topological}. 
We set the lattice constant of the square lattice as unity. 
The model Hamiltonian is given by 
\begin{align}
\label{eq: Ham}
\mathcal{H}=  &2\sum_{\nu=1,2}
\left(-J \Gamma_\nu
+\frac{K}{N} \Gamma_\nu^2
\right)
\nonumber \\
&+
2\sum_{\nu=3,4}
\left(-J' \Gamma_\nu
+\frac{K'}{N} \Gamma_\nu^2
\right)
-H \sum_i S^x_i,
\end{align}
with 
\begin{align}
\label{eq: Gamma}
\Gamma_\nu =\sum_{\alpha=x,y}  S^\alpha_{\bm{Q}_{\nu}}  S^\alpha_{-\bm{Q}_{\nu}}+ (1-I^{\rm EPA}) S^z_{\bm{Q}_{\nu}}  S^z_{-\bm{Q}_{\nu}}.
\end{align}
The first term in Eq.~(\ref{eq: Ham}) represents the interaction at the wave vector $\bm{Q}_1=(\pi/5, 0)$ and $\bm{Q}_2=(0, \pi/5)$, where $\bm{S}_{\bm{Q}_{\nu}}=(S^x_{\bm{Q}_{\nu}}, S^y_{\bm{Q}_{\nu}}, S^z_{\bm{Q}_{\nu}})$ in $\Gamma_\nu$ corresponds to the spin moment at wave vector $\bm{Q}_\nu$; $\bm{S}_{\bm{Q}_{\nu}}$ is related to the classical localized spin $\bm{S}_i=(S^x_i, S^y_i, S^z_i)$ with fixed length $|\bm{S}_i|=1$ via the Fourier transformation. 
$N$ represents the number of sites in the system. 
The first term consists of the bilinear exchange interaction with the coupling constant $J>0$ and the positive biquadratic interaction with the coupling constant $K>0$, the latter of which often appears when the Fermi surface is strongly nested by $\bm{Q}_\nu$ in the itinerant electron systems~\cite{Akagi_PhysRevLett.108.096401, Hayami_PhysRevB.90.060402, Hayami_PhysRevB.95.224424}. 
The factor 2 stands for the contributions from $-\bm{Q}_1$ and $-\bm{Q}_2$. 
The interaction $J$ tends to favor the single-$Q$ spiral state with $\bm{Q}_1$ or $\bm{Q}_2$, while the interaction $K$ tends to favor the multiple-$Q$ state composed of the $\bm{Q}_1$ and $\bm{Q}_2$ modulations~\cite{hayami2021topological}. 
In addition, we introduce the effect of the easy-plane two-spin anisotropic interaction by setting the different form factors for the $xy$ and $z$ spin components, as shown by $\Gamma_\nu$ in Eq.~(\ref{eq: Gamma}); $I^{\rm EPA}>0$ denotes the degree of easy-plane magnetic anisotropy. 
We do not consider the effect of the DM interaction, which often leads to the SkX~\cite{Yi_PhysRevB.80.054416, Hayami_PhysRevLett.121.137202}, owing to the centrosymmetric lattice structure.

The second term in Eq.~(\ref{eq: Ham}) represents the momentum-resolved interaction at the wave vector $\bm{Q}_3=(\pi/5, \pi/5)$ and $\bm{Q}_4=(-\pi/5, \pi/5)$. 
We set the different interaction constants $J'$ and $K'$ for the bilinear and biquadratic exchange interactions, respectively. 
The third term represents the effect of an external in-plane magnetic field along the $x$ direction, where $H$ represents the magnitude of the magnetic field. 

In the following calculations, we parametrize the model parameters as $J=1$, $K=0.2$, $J'=0.6$, and $K'=(J'/J)^2 K$, whereas we change $I^{\rm EPA}$ and $H$; $J$ is the energy unit of the spin model in Eq.~(\ref{eq: Ham}).
The relation $J>J'$ means that the interactions at $\bm{Q}_1$ and $\bm{Q}_2$ are dominant, and those at $\bm{Q}_3$ and $\bm{Q}_4$ give the energy correction. 
Especially, the latter interaction $J'$ plays an important role in inducing the multiple-$Q$ states, since $\bm{Q}_3$ and $\bm{Q}_4$ correspond to the high-harmonic wave vectors of $\bm{Q}_1$ and $\bm{Q}_2$, i.e., $\bm{Q}_3=\bm{Q}_1+\bm{Q}_2$ and $\bm{Q}_4=-\bm{Q}_1+\bm{Q}_2$~\cite{Hayami_PhysRevB.105.174437, hayami2022multiple, hayami2023widely}. 
As will be discussed in Sec.~\ref{sec: Results}, both $K$ and $J'$ are important to stabilize the SkX under the in-plane magnetic field.

\subsection{Method}

We calculate the stable spin configurations of the model in Eq.~(\ref{eq: Ham}) by performing the numerical simulated annealing following the manner in Ref.~\cite{Hayami_PhysRevB.95.224424}. 
First, we start the simulations from a random spin configuration at a high temperature $T_0=1.5$.
Then, we gradually reduce the temperature by the ratio of $T_{n+1}=\alpha T_n$ with $\alpha=0.999999$ up to the final temperature $T=10^{-5}$, where $T_n$ is the $n$th temperature. 
We update all the spins one by one in real space based on the single-spin-flip Metropolis algorithm at each temperature. 
When we reach the final temperature, we perform $10^5$-$10^6$ Monte Carlo sweeps for thermalization and measurements. 
We perform the above procedure while changing $I^{\rm EPA}$ and $H$. 
When we optimize the spin configuration near the phase boundary, we also start the simulations from the spin patterns obtained at low temperatures. 
We set the system size with $N=10^2$ under the periodic boundary condition. 
We confirm that the following results are not altered for larger system sizes, such as  $N=60^2$. 

\subsection{Physical quantities}

We calculate the $\bm{Q}_\nu$ component of magnetic moments, which is given by 
\begin{align}
m^\eta_{\bm{Q}_\nu} = \sqrt{\frac{S^{\eta\eta}_s(\bm{Q}_\nu)}{N}}, 
\end{align}
where $\eta=x,y,z$ and $S^{\eta\eta}_s(\bm{Q}_\nu)$ is the spin structure factor defined by 
\begin{align}
S^{\eta\eta}_s(\bm{Q}_\nu)= \frac{1}{N} \sum_{i,j}S_i^{\eta} S_j^{\eta} e^{i \bm{Q}_\nu \cdot (\bm{r}_i -\bm{r}_j)}. 
\end{align}
Here, $\bm{r}_i$ is the position vector at site $i$ on the square lattice.  
The net magnetization is given by 
\begin{align}
M^\eta = \frac{1}{N} \sum_i S_i^\eta. 
\end{align}

We also calculate the scalar chirality, which is given by 
\begin{align}
\chi^{\rm sc}&= \frac{1}{2 N} 
\sum_{i}
\sum_{\delta,\delta'= \pm1}
\delta \delta'
 \bm{S}_{i} \cdot (\bm{S}_{i+\delta\hat{x}} \times \bm{S}_{i+\delta'\hat{y}}), 
\end{align}
where $\hat{x}$ ($\hat{y}$) represents a shift by lattice constant in the $x$ ($y$) direction on the square lattice.  
The nonzero $\chi^{\rm sc}$ indicates the emergence of the topological spin textures, which includes the SkX.

\section{Results}
\label{sec: Results}

\begin{figure}[tb!]
\begin{center}
\includegraphics[width=1.0\hsize]{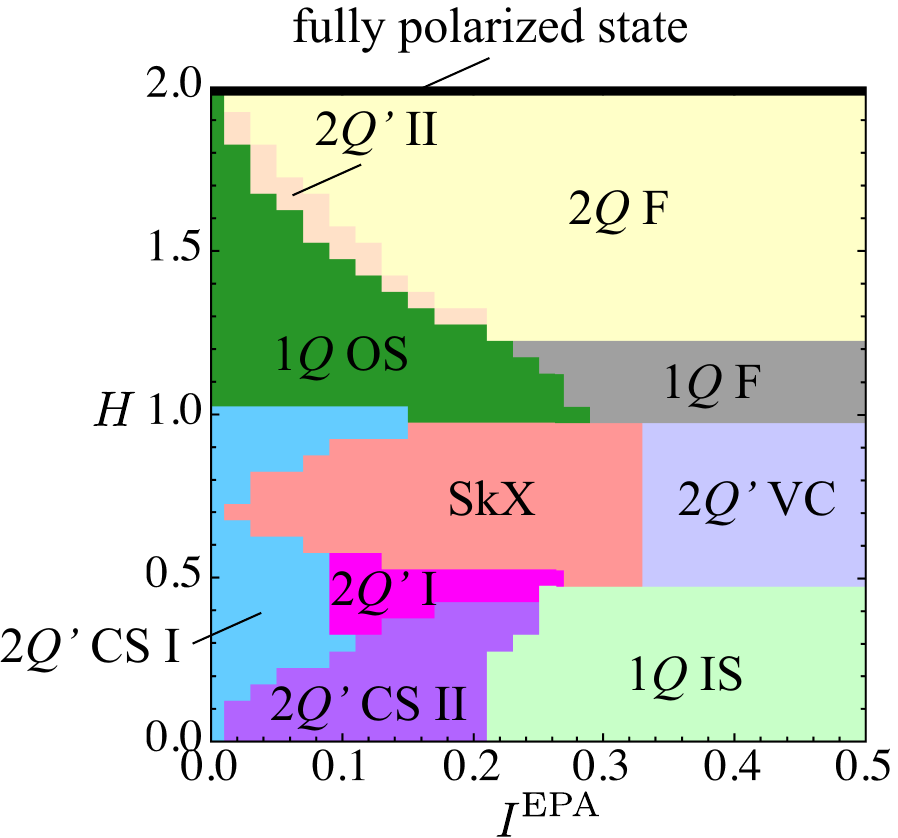} 
\caption{
\label{fig: PD} 
Magnetic phase diagram of the model in Eq.~(\ref{eq: Ham}) on the square lattice by changing the easy-plane anisotropy $I^{\rm EPA}$ and the in-plane magnetic field $H$, which is obtained by the simulated annealing at $K=0.2$ and $J'=0.6$. 
1$Q$ denotes the single-$Q$ states, and 2$Q'$ and $2Q$ represent the anisotropic and isotropic double-$Q$ states, respectively.   
IS, OS, CS, F, VC, and SkX stand for the in-plane spiral, out-of-plane spiral, chiral stripe, fan, vortex crystal, and skyrmion crystal, respectively. 
}
\end{center}
\end{figure}

\begin{figure}[tb!]
\begin{center}
\includegraphics[width=1.0\hsize]{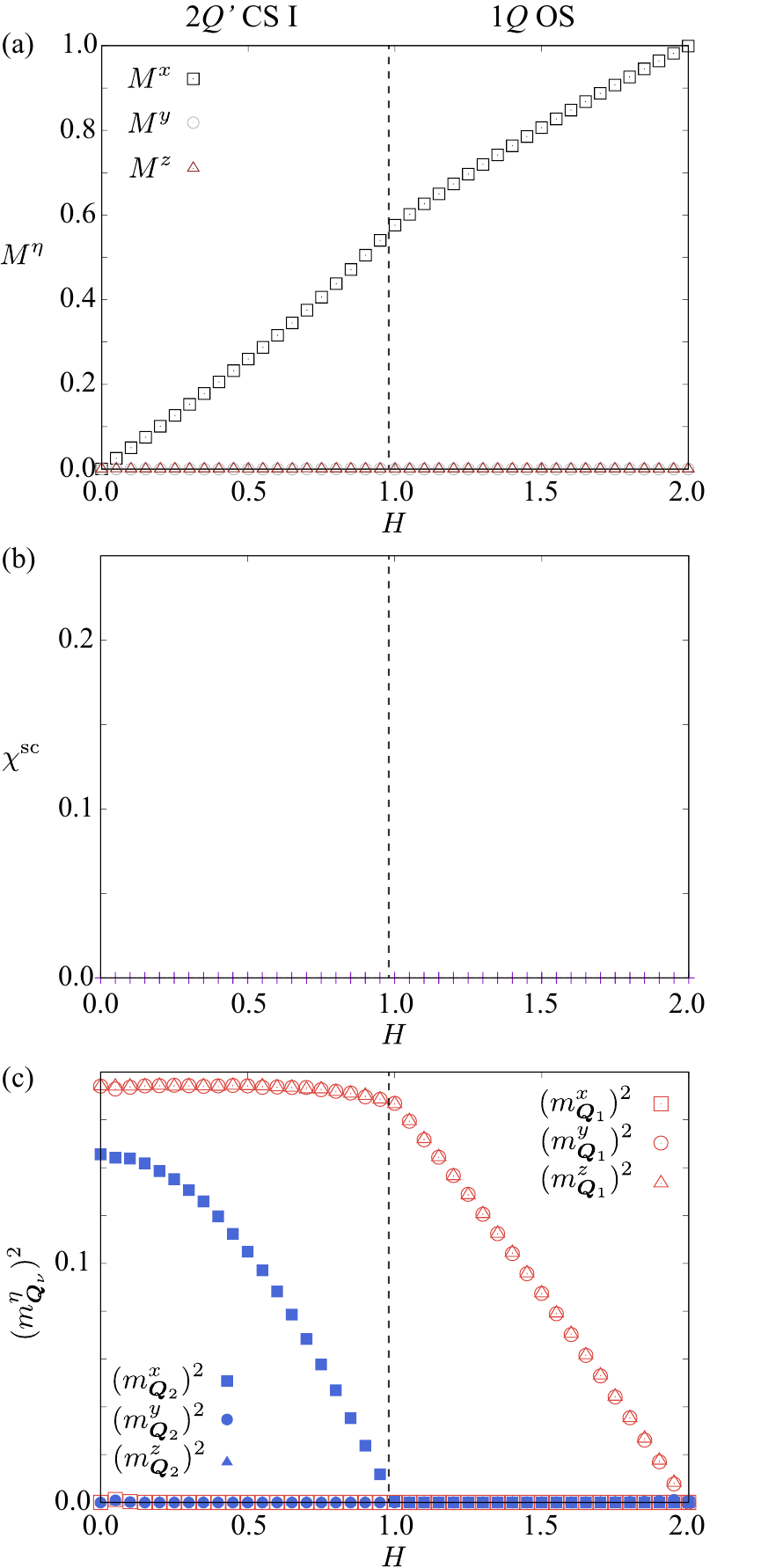} 
\caption{
\label{fig: mag_x=0.00} 
$H$ dependence of (a) the magnetization $M^\eta$, (b) the scalar chirality $\chi^{\rm sc}$, and (c) $\bm{Q}_1$ and $\bm{Q}_2$ components of squared magnetic moments $(m^{\eta}_{\bm{Q}_{1,2}})^2$ for $\eta=x,y,z$ at $I^{\rm EPA}=0$. 
The vertical dashed line represents the phase boundary between different spin states; the corresponding phases are presented above in (a).
}
\end{center}
\end{figure}

\begin{figure}[tb!]
\begin{center}
\includegraphics[width=1.0\hsize]{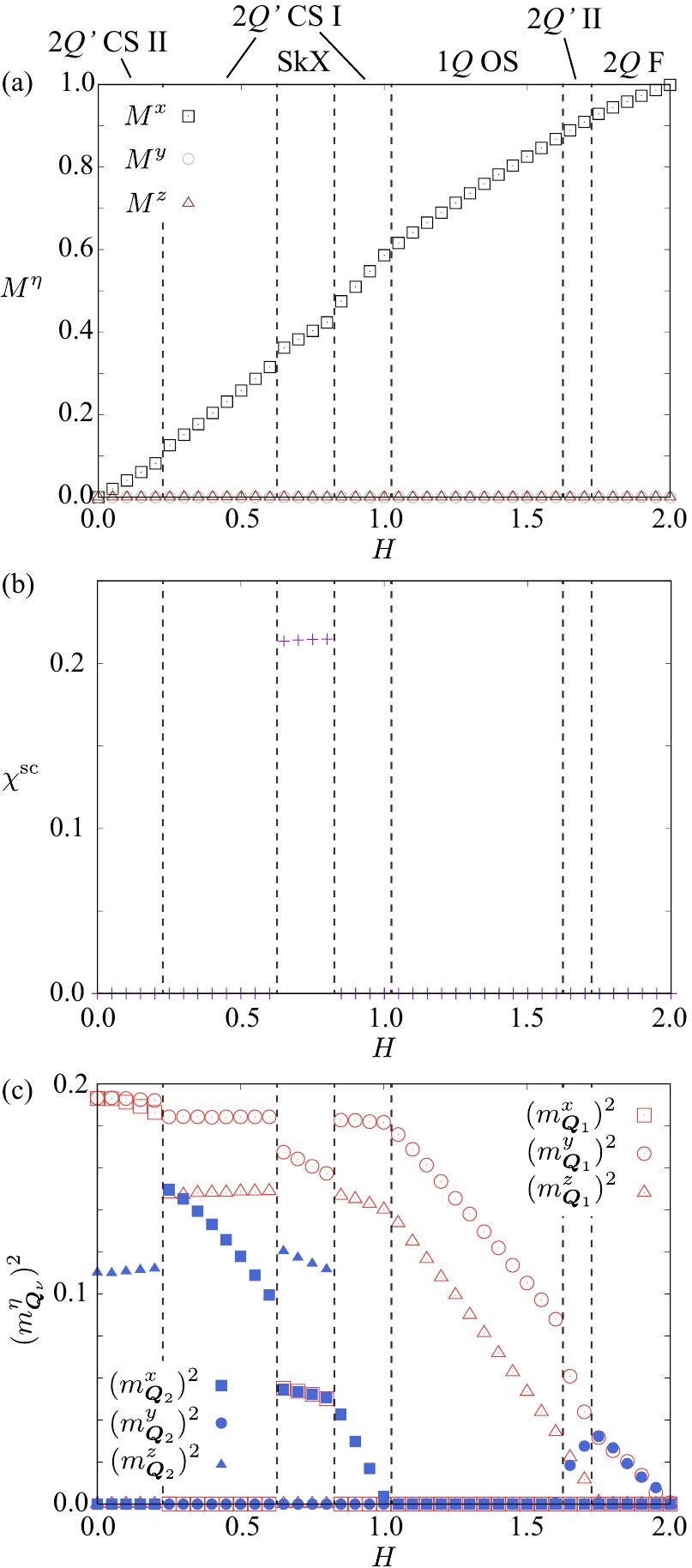} 
\caption{
\label{fig: mag_x=0.06} 
The same plot as in Fig.~\ref{fig: mag_x=0.00} for $I^{\rm EPA}=0.06$. 
}
\end{center}
\end{figure}

\begin{figure}[tb!]
\begin{center}
\includegraphics[width=1.0\hsize]{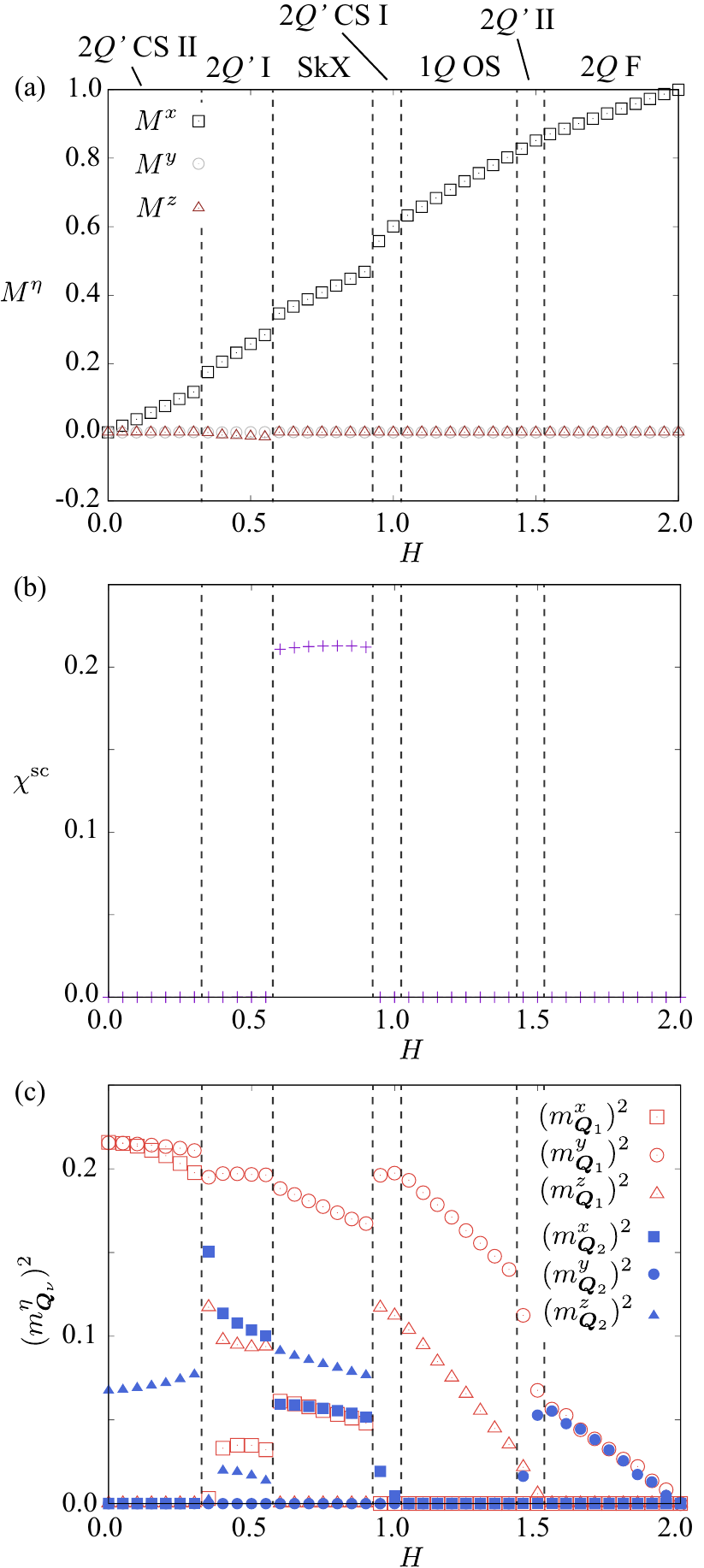} 
\caption{
\label{fig: mag_x=0.12} 
The same plot as in Fig.~\ref{fig: mag_x=0.00} for $I^{\rm EPA}=0.12$. 
}
\end{center}
\end{figure}

\begin{figure}[tb!]
\begin{center}
\includegraphics[width=1.0\hsize]{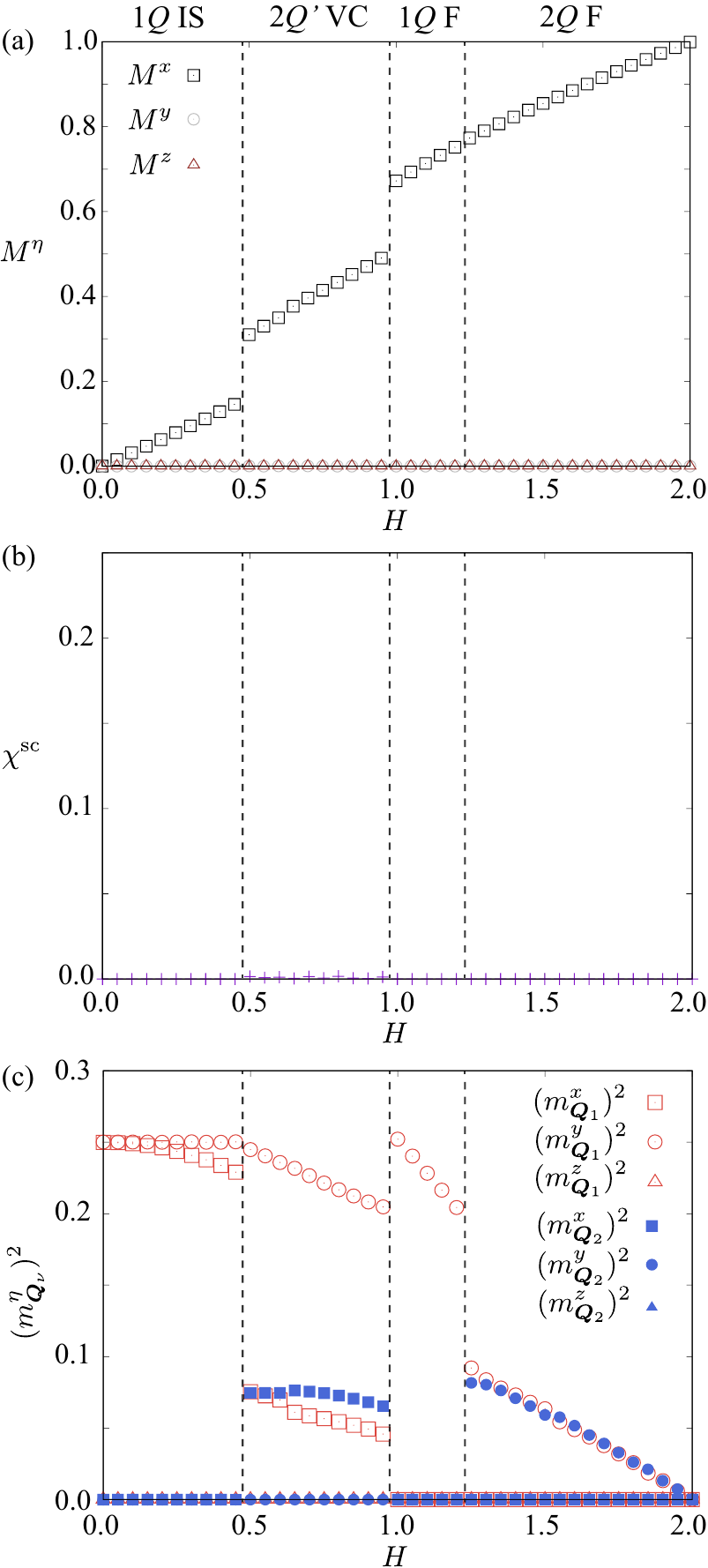} 
\caption{
\label{fig: mag_x=0.40} 
The same plot as in Fig.~\ref{fig: mag_x=0.00} for $I^{\rm EPA}=0.4$. 
}
\end{center}
\end{figure}

By performing the simulated annealing for the model in Eq.~(\ref{eq: Ham}), we obtain the magnetic phase diagram in the plane of $I^{\rm EPA}$ and $H$ at low temperature in Fig.~\ref{fig: PD}. 
Various phases appear depending on the model parameters; the SkX appears in the intermediate $I^{\rm EPA}$ and $H$. 
In the following, we distinguish the obtained states by the $\bm{Q}_1$ and $\bm{Q}_2$ components of the magnetic moments without referring to the $\bm{Q}_3$ and $\bm{Q}_4$ components, since the former amplitude is larger than the latter one; we term a state with nonzero $(m^{\eta}_{\bm{Q}_{1}})^2$ or $(m^{\eta}_{\bm{Q}_{2}})^2$ as a single-$Q$ (1$Q$) state and that with both nonzero $(m^{\eta}_{\bm{Q}_{1}})^2$ and $(m^{\eta}_{\bm{Q}_{2}})^2$ as a double-$Q$ (2$Q$) state. 
We present the data in terms of the uniform magnetization, scalar chirality, and the $\bm{Q}_1$ and $\bm{Q}_2$ components of the magnetic moments in Figs.~\ref{fig: mag_x=0.00}--\ref{fig: mag_x=0.40} for different $I^{\rm EPA}$. 

\begin{figure}[tb!]
\begin{center}
\includegraphics[width=0.93\hsize]{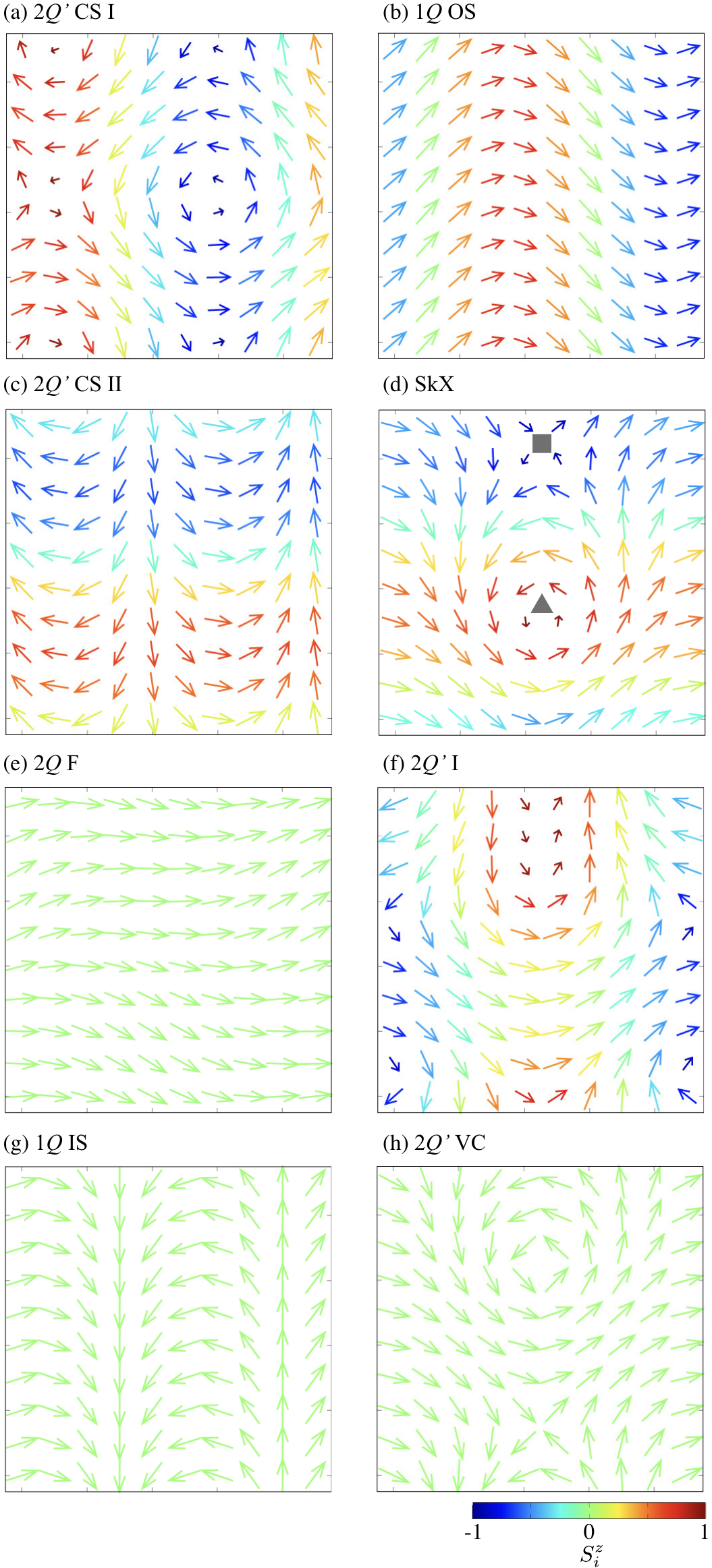} 
\caption{
\label{fig: Spin} 
Real-space spin configurations in 
(a) the 2$Q'$ CS I state at $I^{\rm EPA}=0$ and $H=0.05$, 
(b) the 1$Q$ OS state at $I^{\rm EPA}=0$ and $H=1.2$, 
(c) the 2$Q'$ CS II state at $I^{\rm EPA}=0.06$ and $H=0$, 
(d) the SkX at $I^{\rm EPA}=0.06$ and $H=0.7$, 
(e) the 2$Q$ F state at $I^{\rm EPA}=0.06$ and $H=1.85$, 
(f) the 2$Q'$ I state at $I^{\rm EPA}=0.12$ and $H=0.5$, 
(g) the 1$Q$ IS state at $I^{\rm EPA}=0.4$ and $H=0$, and 
(h) the 2$Q'$ VC state at $I^{\rm EPA}=0.4$ and $H=0.75$. 
The arrows represent the direction of the in-plane spin and the color shows its $z$ component. 
In (d), the triangle and square stand for the vortex and antivortex, respectively. 
}
\end{center}
\end{figure}

\begin{figure}[tb!]
\begin{center}
\includegraphics[width=1.0\hsize]{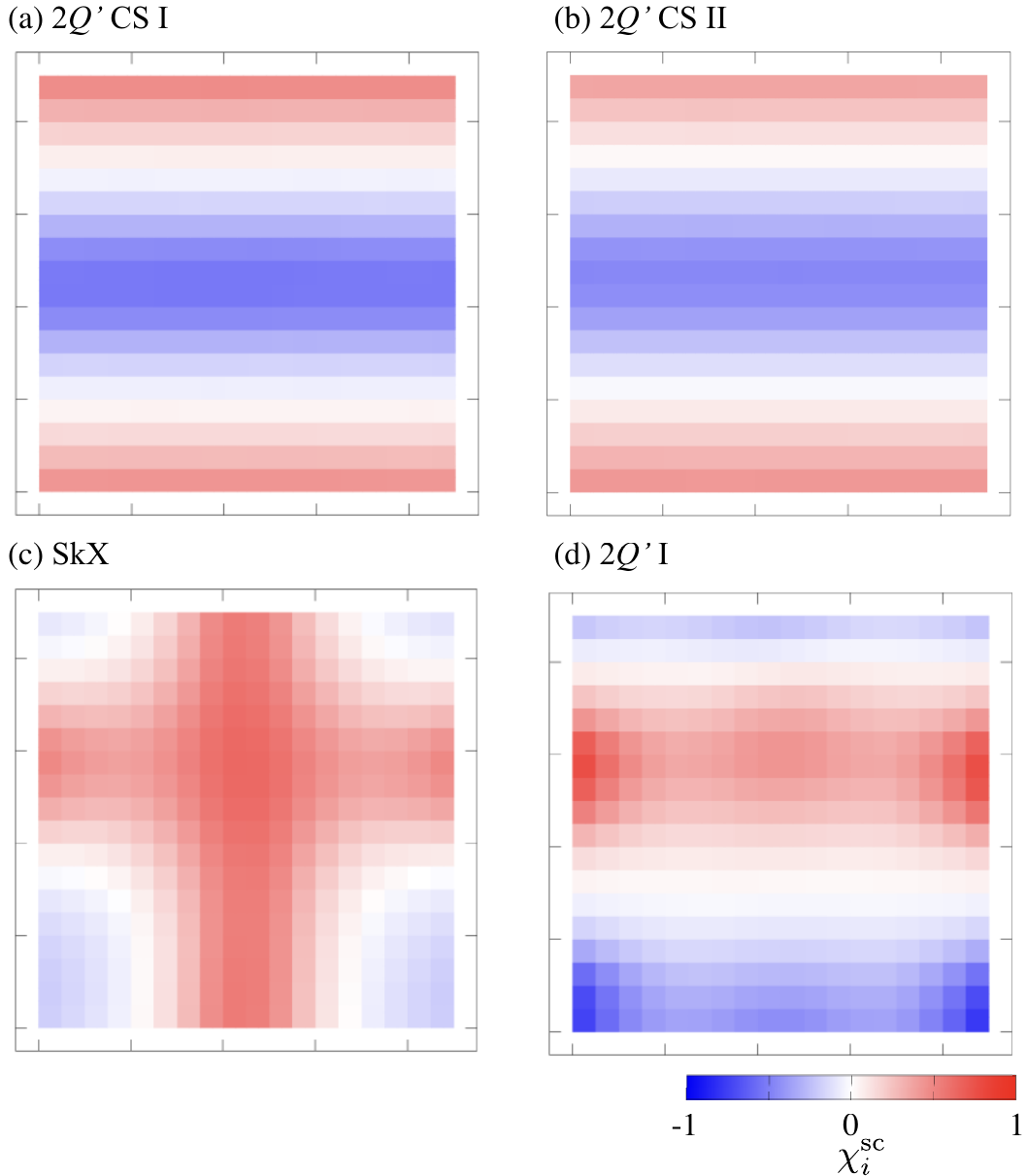} 
\caption{
\label{fig: chirality} 
Real-space scalar chirality configurations in 
(a) the 2$Q'$ CS I state at $I^{\rm EPA}=0$ and $H=0.05$, 
(b) the 2$Q'$ CS II state at $I^{\rm EPA}=0.06$ and $H=0$, 
(c) the SkX at $I^{\rm EPA}=0.06$ and $H=0.7$, and 
(d) the 2$Q'$ I state at $I^{\rm EPA}=0.12$ and $H=0.5$. 
}
\end{center}
\end{figure}

For $I^{\rm EPA}=0$, the zero-field to low-field state corresponds to the 2$Q'$ CS I state, where CS stands for the chiral stripe and $Q'$ means the different amplitudes of $(m^{x}_{\bm{Q}_{1}})^2 + (m^{y}_{\bm{Q}_{1}})^2 + (m^{z}_{\bm{Q}_{1}})^2$ and $(m^{x}_{\bm{Q}_{2}})^2+(m^{y}_{\bm{Q}_{2}})^2+(m^{z}_{\bm{Q}_{2}})^2$. 
The spin configuration in this state is characterized by $(m^{y}_{\bm{Q}_{1}})^2$ and $(m^{z}_{\bm{Q}_{1}})^2$ as the dominant contribution and $(m^{x}_{\bm{Q}_{2}})^2$ as the subdominant contribution, as shown in Fig.~\ref{fig: mag_x=0.00}(c). 
In other words, this spin configuration is represented by a superposition of the out-of-plane spiral wave at the $\bm{Q}_1$ component and the sinusoidal wave at the $\bm{Q}_2$ component. 
The orthonormal relation of the spin oscillations between the $\bm{Q}_1$ and $\bm{Q}_2$ components is due to preventing the energy loss under the spin length constraint with $|\bm{S}_i|=1$~\cite{Solenov_PhysRevLett.108.096403}. 
Reflecting the double-$Q$ superposition in different spin components, the 2$Q'$ CS I state exhibits the noncoplanar spin texture, as shown by the real-space spin configuration in Fig.~\ref{fig: Spin}(a), although they do not show a net scalar chirality in Fig.~\ref{fig: mag_x=0.00}(b). 
Instead, this state accompanies the chirality density wave along the $\bm{Q}_2$ direction, as shown in Fig.~\ref{fig: chirality}(a)~\cite{Hayami_PhysRevB.94.174420}. 

The emergence of the 2$Q'$ CS I state at $I^{\rm EPA}=0$ is owing to the positive biquadratic interaction $K$ rather than $J'$; $K$ tends to enhance the stability of multiple-$Q$ states~\cite{Hayami_PhysRevB.95.224424}.  
Indeed, this state remains stable when $J'=0$, as will be discussed in Appendix~\ref{sec: Results for other model parameters}.

As $H$ increases, $M^x$ gradually increases and $(m^{x}_{\bm{Q}_{2}})^2$ decreases, as shown in Figs.~\ref{fig: mag_x=0.00}(a) and \ref{fig: mag_x=0.00}(c), respectively.  
Then, the 2$Q'$ CS I state shows the phase transition to the 1$Q$ OS state at $H \simeq 0.98$. 
The 1$Q$ OS state corresponds to the single-$Q$ out-of-plane spiral state, whose spiral plane lies in the $yz$ plane in order to gain the Zeeman energy; it is noted that the spiral plane is always fixed in the $yz$ plane irrespective of $\bm{Q}_{1,2}$ owing to the absence of the $\bm{Q}$-dependent anisotropy. 
The real-space spin configuration of the 1$Q$ OS state is shown in Fig.~\ref{fig: Spin}(b). 
The 1$Q$ OS state turns into the fully polarized state at $H=2$. 

When the effect of $I^{\rm EPA}$ is considered, the 2$Q'$ CS I state at zero field is replaced by the 2$Q'$ CS II state, as shown in Fig.~\ref{fig: PD}. 
We show the $H$ dependence of physical quantities at $I^{\rm EPA}=0.06$ in Fig.~\ref{fig: mag_x=0.06}. 
The spin texture of the 2$Q'$ CS II state is characterized by $(m^{x}_{\bm{Q}_{1}})^2$ and $(m^{y}_{\bm{Q}_{1}})^2$ as the dominant contribution and $(m^{z}_{\bm{Q}_{2}})^2$ as the subdominant contribution, as shown in Fig.~\ref{fig: mag_x=0.06}(c). 
In contrast to the 2$Q'$ CS I state, the spiral plane changes from the $yz$ plane to the $xy$ plane in order to gain the energy by $I^{\rm EPA}$. 
Accordingly, the sinusoidal modulation occurs for the $z$ spin component along the $\bm{Q}_2$ direction. 
The real-space spin and scalar chirality configurations are shown in Figs.~\ref{fig: Spin}(c) and \ref{fig: chirality}(b), respectively, where the chirality density wave occurs along the $\bm{Q}_2$ direction similar to the 2$Q'$ CS I state. 
It is noted that the uniform component of the scalar chirality is zero in the 2$Q'$ CS II state [Fig.~\ref{fig: mag_x=0.06}(b)]. 

The 2$Q'$ CS II state changes into the 2$Q'$ CS I state with jumps of $M^x$ and $m^\eta_{\bm{Q}_\nu}$ at $H \simeq 0.23$, as shown in Figs.~\ref{fig: mag_x=0.06}(a) and \ref{fig: mag_x=0.06}(c). 
This phase transition is regarded as the spin-flop phase transition, where the spiral plane changes from the $xy$ plane parallel to the field direction to the $yz$ plane perpendicular to the field direction. 

While further increasing $H$, we find that the SkX is stabilized for $0.63 \lesssim H \lesssim 0.83$, which is characterized by nonzero scalar chirality $\chi^{\rm sc}$ in Fig.~\ref{fig: mag_x=0.06}(b).  
The spin configuration of the SkX is expressed as the linear combination of the spiral wave on the $xy$ plane along the $\bm{Q}_1$ direction and that on the $yz$ plane along the $\bm{Q}_2$ direction, as shown in Fig.~\ref{fig: mag_x=0.06}(c), where the amplitudes of the magnetic moments at the $\bm{Q}_1$ and $\bm{Q}_2$ components are different owing to the lack of the fourfold rotational symmetry under the in-plane field; this feature is different from the SkX stabilized in the out-of-plane field, where the fourfold rotational spin texture appears. 

As shown by the real-space spin configuration in Fig.~\ref{fig: Spin}(d), the SkX phase consists of two vortices: One is the vortex with the positive $z$-spin polarization, which is denoted as the triangle, and the other is the antivortex with the negative $z$-spin polarization, which is denoted as the square. 
Since both the winding number and $z$-spin polarization are opposite between the vortex and antivortex, these vortices lead to the same sign of the scalar chirality, as shown in Fig.~\ref{fig: chirality}(c). 
Although the SkX in Fig.~\ref{fig: Spin}(d) exhibits the positive scalar chirality, the state with the negative scalar chirality is degenerate; the degeneracy between them is lifted when the effect of bond-dependent magnetic anisotropy in $\bm{Q}_1$--$\bm{Q}_4$ is taken into account~\cite{Hayami_doi:10.7566/JPSJ.89.103702, Hayami_PhysRevB.105.104428}. 
The skyrmion core with $S_i^x = -1$ is found between the vortex and antivortex, which is located at the interstitial site rather than the lattice site~\cite{Hayami_PhysRevResearch.3.043158}.

From the energetic viewpoint, the SkX is stabilized by the interplay among $K$, $J'$, $I^{\rm EPA}$, and $H$ in the model Hamiltonian in Eq.~(\ref{eq: Ham}). 
Among them, we find that $J'$, $I^{\rm EPA}$, and $H$ play an important role in inducing the SkX, while $K$ enhances the stability of the SkX region. 
In other words, the SkX remains stable for $K=0$, while it disappears for $J'=0$, as shown in Appendix~\ref{sec: Results for other model parameters}. 
This means that the SkX can be realized in both itinerant magnets and insulating magnets, where the contribution from $K$ is negligible for the latter. 
Thus, the application of the in-plane external magnetic field to materials with easy-plane magnetic anisotropy is one of the possibilities to induce the SkX in centrosymmetric systems irrespective of metals and insulators. 

The SkX turns into the 2$Q'$ CS I state again by increasing $H$. 
Then, the 2$Q'$ CS I state changes into the 1$Q$ OS state similar to the result at $I^{\rm EPA}=0$.  
The increase of $H$ in the 1$Q$ OS state leads to two additional phases before entering into the fully polarized state: One is the 2$Q'$ II state and the other is the 2$Q$ F state, where F means the fan. 
In the 2$Q'$ II state, $(m^y_{\bm{Q}_2})^2$ becomes nonzero in addition to $(m^y_{\bm{Q}_1})^2$ and $(m^z_{\bm{Q}_1})^2$. 
Meanwhile, the 2$Q$ F state is characterized by the double-$Q$ structure in the $y$ component with the same intensity, i.e. $(m^y_{\bm{Q}_1})^2=(m^y_{\bm{Q}_2})^2$, as shown in Fig.~\ref{fig: mag_x=0.06}(c), where the spin configuration is shown in Fig.~\ref{fig: Spin}(e). 

By increasing $I^{\rm EPA}$, the regions of $2Q'$ CS I and 1$Q$ OS phases become narrower, while those of the SkX, 2$Q'$ CS II, and 2$Q$ F phases become wider, as shown in Fig.~\ref{fig: PD}. 
For $I^{\rm EPA} \gtrsim 0.1$, the 2$Q'$ I state appears between the SkX and 2$Q'$ CS II state; the phase sequence in the case of $I^{\rm EPA}=0.12$ is shown in Fig.~\ref{fig: mag_x=0.12}. 
The 2$Q'$ I state is characterized by nonzero $(m^x_{\bm{Q}_1})^2$, $(m^y_{\bm{Q}_1})^2$, $(m^z_{\bm{Q}_1})^2$, $(m^x_{\bm{Q}_2})^2$, and $(m^z_{\bm{Q}_2})^2$, as shown in Fig.~\ref{fig: mag_x=0.12}(c); the spiral plane in the $\bm{Q}_1$ component lies between the $xy$ and $yz$ plane. 
Accordingly, this state shows nonzero uniform magnetization in the $z$ component, $M^z$, as shown in Fig.~\ref{fig: mag_x=0.12}(a). 
The real-space spin configuration of the 2$Q'$ I state is shown in Fig.~\ref{fig: Spin}(f). 
Although this state accompanies the noncoplanar spin texture similar to the SkX, the scalar chirality becomes zero in the whole system in Fig.~\ref{fig: mag_x=0.12}(b); only the chirality density wave is found, as shown in Fig.~\ref{fig: chirality}(d).

When $I^{\rm EPA}$ further increases, the SkX vanishes around $I^{\rm EPA} \simeq 0.33$, and all the phases are characterized by the coplanar spin textures in the $xy$ plane. 
In the case of $I^{\rm EPA}=0.4$, the 1$Q$ IS state is stabilized in the low-field region instead of the 2$Q'$ CS II state; the spin configuration in the former state is obtained by eliminating the $\bm{Q}_2$ component in the $z$ spin in the latter state. 
Since $I^{\rm EPA}$ favors the in-plane spin texture, the phase transition from the 2$Q'$ CS II state to the 1$Q$ IS state in increasing $I^{\rm EPA}$ is reasonable.  
The spin configuration in real space is shown in Fig.~\ref{fig: Spin}(g). 

The increase of $H$ for $I^{\rm EPA}=0.4$ leads to the phase transition from the 1$Q$ IS state to the 2$Q'$ VC state with jumps of $M^x$ and $m^\eta_{\bm{Q}_\nu}$, as shown in Figs.~\ref{fig: mag_x=0.40}(a) and \ref{fig: mag_x=0.40}(c); VC means the vortex crystal. 
The 2$Q'$ VC state is characterized by the spiral wave in the $xy$ plane along the $\bm{Q}_1$ direction and the sinusoidal wave in the $x$ component along the $\bm{Q}_2$ direction. 
Owing to the double-$Q$ superposition in terms of the $x$-spin component, the real-space spin configuration consists of the vortex and antivortex, as shown in Fig.~\ref{fig: Spin}(h), which is similar to that in the SkX.  
Similar coplanar spin textures accompanying the vortex and antivortex have been found in triangular magnets with the easy-plane magnetic anisotropy~\cite{hayami2020multiple}, such as Y$_3$Co$_8$Sn$_4$~\cite{takagi2018multiple}. 
Our results indicate that the coplanar vortex crystal can be realized even in the centrosymmetric tetragonal host when the in-plane magnetic field is applied.

\section{Summary}
\label{sec: Summary}

To summarize, we have investigated the stability of the SkX in centrosymmetric tetragonal magnets with easy-plane magnetic anisotropy by applying the in-plane external magnetic field. 
The data were obtained by performing the simulated annealing for the spin model including the biquadratic interaction, high-harmonic wave-vector interaction, and easy-plane two-spin anisotropic interaction. 
As a result, we have found that their synergy tends to favor the SkX in the in-plane field. 
The SkX is stabilized in the intermediate-field region when the strength of easy-plane anisotropy is moderate. 
We have also found that the high-harmonic wave-vector interaction plays an important role in inducing the SkX, while the biquadratic interaction enhances its stability. 
Our results suggest the importance of the applied field direction; the in-plane magnetic field is essential to realize the SkX in centrosymmetric magnets with easy-plane magnetic anisotropy.

\begin{acknowledgments}
This research was supported by JSPS KAKENHI Grants Numbers JP21H01037, JP22H04468, JP22H00101, JP22H01183, JP23H04869, JP23K03288, and by JST PRESTO (JPMJPR20L8) and JST CREST (JPMJCR23O4).  
Parts of the numerical calculations were performed in the supercomputing systems in ISSP, the University of Tokyo.
\end{acknowledgments}

\appendix

\section{Phase diagrams for other model parameters}
\label{sec: Results for other model parameters}

\begin{figure}[tb!]
\begin{center}
\includegraphics[width=1.0\hsize]{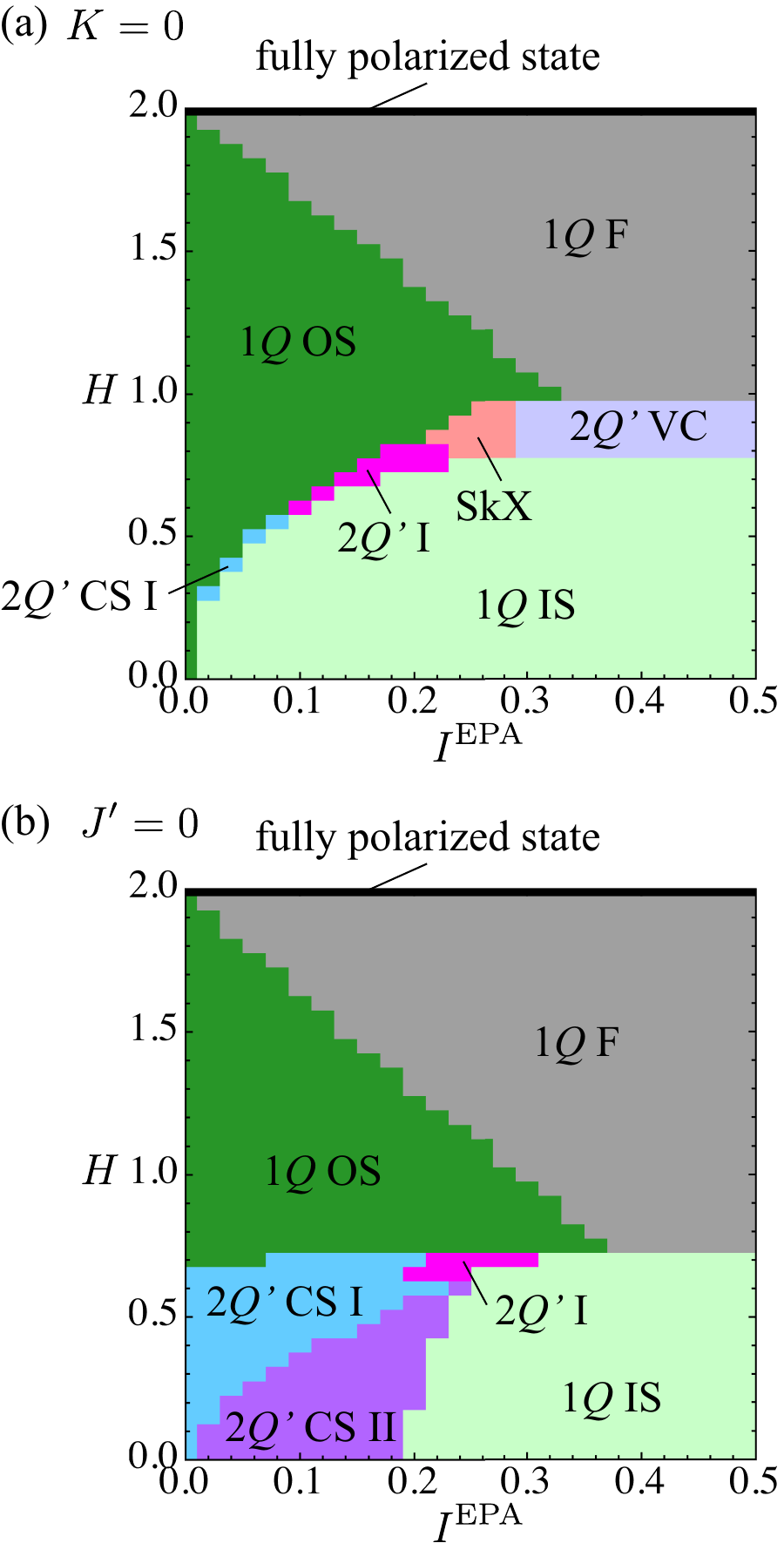} 
\caption{
\label{fig: PD_other} 
Magnetic phase diagrams by setting (a) $K=0$ and (b) $J'=0$. 
The other model parameters are the same as those used in Fig.~\ref{fig: PD}. 
}
\end{center}
\end{figure}

We briefly discuss the magnetic phase diagrams when the effect of the biquadratic interaction $K$ or high-harmonic wave-vector interaction $J'$ is ignored. 
Figure~\ref{fig: PD_other}(a) shows the phase diagram in the case of $K=0$. 
As shown in the phase diagram, the SkX is stabilized for intermediate $I^{\rm EPA}$ and $H$, although its stability region becomes narrower compared to the phase diagram at $K=0.2$ in Fig.~\ref{fig: PD}.
This result indicates that the high-harmonic wave-vector interaction $J'$ is significant in stabilizing the SkX in Fig.~\ref{fig: PD}, but its stability region is narrow. 
Meanwhile, some of the other phases, $2Q'$ CS II, 2$Q'$ II, and 2$Q$ F, vanish in Fig.~\ref{fig: PD_other}(a), which means that solely $J'$ is not enough to induce these phases. 

Next, Fig.~\ref{fig: PD_other}(b) shows the phase diagram in the case of $J'=0$. 
In contrast to the phase diagram in Fig.~\ref{fig: PD_other}(a), the SkX and 2$Q'$ VC state do not appear in the phase diagram. 
In other words, the biquadratic interaction $K$ is not enough to induce these two phases within the present model parameters.

\bibliographystyle{apsrev}
\bibliography{../ref.bib}

\begin{thebibliography}{55}
\expandafter\ifx\csname natexlab\endcsname\relax\def\natexlab#1{#1}\fi
\expandafter\ifx\csname bibnamefont\endcsname\relax
  \def\bibnamefont#1{#1}\fi
\expandafter\ifx\csname bibfnamefont\endcsname\relax
  \def\bibfnamefont#1{#1}\fi
\expandafter\ifx\csname citenamefont\endcsname\relax
  \def\citenamefont#1{#1}\fi
\expandafter\ifx\csname url\endcsname\relax
  \def\url#1{\texttt{#1}}\fi
\expandafter\ifx\csname urlprefix\endcsname\relax\def\urlprefix{URL }\fi
\providecommand{\bibinfo}[2]{#2}
\providecommand{\eprint}[2][]{\url{#2}}

\bibitem[{\citenamefont{Bogdanov and Hubert}(1994{\natexlab{a}})}]{Bogdanov94}
\bibinfo{author}{\bibfnamefont{A.}~\bibnamefont{Bogdanov}} \bibnamefont{and}
  \bibinfo{author}{\bibfnamefont{A.}~\bibnamefont{Hubert}},
  \bibinfo{journal}{J. Magn. Magn. Mater.} \textbf{\bibinfo{volume}{138}},
  \bibinfo{pages}{255 } (\bibinfo{year}{1994}{\natexlab{a}}), ISSN
  \bibinfo{issn}{0304-8853}.

\bibitem[{\citenamefont{Bogdanov and
  Hubert}(1994{\natexlab{b}})}]{bocdanov1994properties}
\bibinfo{author}{\bibfnamefont{A.}~\bibnamefont{Bogdanov}} \bibnamefont{and}
  \bibinfo{author}{\bibfnamefont{A.}~\bibnamefont{Hubert}},
  \bibinfo{journal}{Phys. Stat. Sol. (b)} \textbf{\bibinfo{volume}{186}},
  \bibinfo{pages}{527} (\bibinfo{year}{1994}{\natexlab{b}}).

\bibitem[{\citenamefont{Bogdanov and Hubert}(1999)}]{bogdanov1999stability}
\bibinfo{author}{\bibfnamefont{A.}~\bibnamefont{Bogdanov}} \bibnamefont{and}
  \bibinfo{author}{\bibfnamefont{A.}~\bibnamefont{Hubert}},
  \bibinfo{journal}{J. Magn. Magn. Mater.} \textbf{\bibinfo{volume}{195}},
  \bibinfo{pages}{182} (\bibinfo{year}{1999}).

\bibitem[{\citenamefont{Nagaosa and Tokura}(2013)}]{nagaosa2013topological}
\bibinfo{author}{\bibfnamefont{N.}~\bibnamefont{Nagaosa}} \bibnamefont{and}
  \bibinfo{author}{\bibfnamefont{Y.}~\bibnamefont{Tokura}},
  \bibinfo{journal}{Nat. Nanotechnol.} \textbf{\bibinfo{volume}{8}},
  \bibinfo{pages}{899} (\bibinfo{year}{2013}).

\bibitem[{\citenamefont{Finocchio et~al.}(2016)\citenamefont{Finocchio,
  B{\"u}ttner, Tomasello, Carpentieri, and Kl{\"a}ui}}]{finocchio2016magnetic}
\bibinfo{author}{\bibfnamefont{G.}~\bibnamefont{Finocchio}},
  \bibinfo{author}{\bibfnamefont{F.}~\bibnamefont{B{\"u}ttner}},
  \bibinfo{author}{\bibfnamefont{R.}~\bibnamefont{Tomasello}},
  \bibinfo{author}{\bibfnamefont{M.}~\bibnamefont{Carpentieri}},
  \bibnamefont{and}
  \bibinfo{author}{\bibfnamefont{M.}~\bibnamefont{Kl{\"a}ui}},
  \bibinfo{journal}{J. Phys. D: Appl. Phys.} \textbf{\bibinfo{volume}{49}},
  \bibinfo{pages}{423001} (\bibinfo{year}{2016}).

\bibitem[{\citenamefont{Fert et~al.}(2017)\citenamefont{Fert, Reyren, and
  Cros}}]{fert2017magnetic}
\bibinfo{author}{\bibfnamefont{A.}~\bibnamefont{Fert}},
  \bibinfo{author}{\bibfnamefont{N.}~\bibnamefont{Reyren}}, \bibnamefont{and}
  \bibinfo{author}{\bibfnamefont{V.}~\bibnamefont{Cros}},
  \bibinfo{journal}{Nat. Rev. Mater.} \textbf{\bibinfo{volume}{2}},
  \bibinfo{pages}{17031} (\bibinfo{year}{2017}).

\bibitem[{\citenamefont{Zhang et~al.}(2020)\citenamefont{Zhang, Zhou, Song,
  Park, Xia, Ezawa, Liu, Zhao, Zhao, and Woo}}]{zhang2020skyrmion}
\bibinfo{author}{\bibfnamefont{X.}~\bibnamefont{Zhang}},
  \bibinfo{author}{\bibfnamefont{Y.}~\bibnamefont{Zhou}},
  \bibinfo{author}{\bibfnamefont{K.~M.} \bibnamefont{Song}},
  \bibinfo{author}{\bibfnamefont{T.-E.} \bibnamefont{Park}},
  \bibinfo{author}{\bibfnamefont{J.}~\bibnamefont{Xia}},
  \bibinfo{author}{\bibfnamefont{M.}~\bibnamefont{Ezawa}},
  \bibinfo{author}{\bibfnamefont{X.}~\bibnamefont{Liu}},
  \bibinfo{author}{\bibfnamefont{W.}~\bibnamefont{Zhao}},
  \bibinfo{author}{\bibfnamefont{G.}~\bibnamefont{Zhao}}, \bibnamefont{and}
  \bibinfo{author}{\bibfnamefont{S.}~\bibnamefont{Woo}}, \bibinfo{journal}{J.
  Phys.: Condens. Matter} \textbf{\bibinfo{volume}{32}},
  \bibinfo{pages}{143001} (\bibinfo{year}{2020}).

\bibitem[{\citenamefont{M{\"u}hlbauer et~al.}(2009)\citenamefont{M{\"u}hlbauer,
  Binz, Jonietz, Pfleiderer, Rosch, Neubauer, Georgii, and
  B{\"o}ni}}]{Muhlbauer_2009skyrmion}
\bibinfo{author}{\bibfnamefont{S.}~\bibnamefont{M{\"u}hlbauer}},
  \bibinfo{author}{\bibfnamefont{B.}~\bibnamefont{Binz}},
  \bibinfo{author}{\bibfnamefont{F.}~\bibnamefont{Jonietz}},
  \bibinfo{author}{\bibfnamefont{C.}~\bibnamefont{Pfleiderer}},
  \bibinfo{author}{\bibfnamefont{A.}~\bibnamefont{Rosch}},
  \bibinfo{author}{\bibfnamefont{A.}~\bibnamefont{Neubauer}},
  \bibinfo{author}{\bibfnamefont{R.}~\bibnamefont{Georgii}}, \bibnamefont{and}
  \bibinfo{author}{\bibfnamefont{P.}~\bibnamefont{B{\"o}ni}},
  \bibinfo{journal}{Science} \textbf{\bibinfo{volume}{323}},
  \bibinfo{pages}{915} (\bibinfo{year}{2009}).

\bibitem[{\citenamefont{Yu et~al.}(2010)\citenamefont{Yu, Onose, Kanazawa,
  Park, Han, Matsui, Nagaosa, and Tokura}}]{yu2010real}
\bibinfo{author}{\bibfnamefont{X.~Z.} \bibnamefont{Yu}},
  \bibinfo{author}{\bibfnamefont{Y.}~\bibnamefont{Onose}},
  \bibinfo{author}{\bibfnamefont{N.}~\bibnamefont{Kanazawa}},
  \bibinfo{author}{\bibfnamefont{J.~H.} \bibnamefont{Park}},
  \bibinfo{author}{\bibfnamefont{J.~H.} \bibnamefont{Han}},
  \bibinfo{author}{\bibfnamefont{Y.}~\bibnamefont{Matsui}},
  \bibinfo{author}{\bibfnamefont{N.}~\bibnamefont{Nagaosa}}, \bibnamefont{and}
  \bibinfo{author}{\bibfnamefont{Y.}~\bibnamefont{Tokura}},
  \bibinfo{journal}{Nature} \textbf{\bibinfo{volume}{465}},
  \bibinfo{pages}{901} (\bibinfo{year}{2010}).

\bibitem[{\citenamefont{Yu et~al.}(2011)\citenamefont{Yu, Kanazawa, Onose,
  Kimoto, Zhang, Ishiwata, Matsui, and Tokura}}]{yu2011near}
\bibinfo{author}{\bibfnamefont{X.~Z.} \bibnamefont{Yu}},
  \bibinfo{author}{\bibfnamefont{N.}~\bibnamefont{Kanazawa}},
  \bibinfo{author}{\bibfnamefont{Y.}~\bibnamefont{Onose}},
  \bibinfo{author}{\bibfnamefont{K.}~\bibnamefont{Kimoto}},
  \bibinfo{author}{\bibfnamefont{W.}~\bibnamefont{Zhang}},
  \bibinfo{author}{\bibfnamefont{S.}~\bibnamefont{Ishiwata}},
  \bibinfo{author}{\bibfnamefont{Y.}~\bibnamefont{Matsui}}, \bibnamefont{and}
  \bibinfo{author}{\bibfnamefont{Y.}~\bibnamefont{Tokura}},
  \bibinfo{journal}{Nat. Mater.} \textbf{\bibinfo{volume}{10}},
  \bibinfo{pages}{106} (\bibinfo{year}{2011}).

\bibitem[{\citenamefont{Seki et~al.}(2012)\citenamefont{Seki, Yu, Ishiwata, and
  Tokura}}]{seki2012observation}
\bibinfo{author}{\bibfnamefont{S.}~\bibnamefont{Seki}},
  \bibinfo{author}{\bibfnamefont{X.~Z.} \bibnamefont{Yu}},
  \bibinfo{author}{\bibfnamefont{S.}~\bibnamefont{Ishiwata}}, \bibnamefont{and}
  \bibinfo{author}{\bibfnamefont{Y.}~\bibnamefont{Tokura}},
  \bibinfo{journal}{Science} \textbf{\bibinfo{volume}{336}},
  \bibinfo{pages}{198} (\bibinfo{year}{2012}).

\bibitem[{\citenamefont{Heinze et~al.}(2011)\citenamefont{Heinze, von Bergmann,
  Menzel, Brede, Kubetzka, Wiesendanger, Bihlmayer, and
  Bl{\"u}gel}}]{heinze2011spontaneous}
\bibinfo{author}{\bibfnamefont{S.}~\bibnamefont{Heinze}},
  \bibinfo{author}{\bibfnamefont{K.}~\bibnamefont{von Bergmann}},
  \bibinfo{author}{\bibfnamefont{M.}~\bibnamefont{Menzel}},
  \bibinfo{author}{\bibfnamefont{J.}~\bibnamefont{Brede}},
  \bibinfo{author}{\bibfnamefont{A.}~\bibnamefont{Kubetzka}},
  \bibinfo{author}{\bibfnamefont{R.}~\bibnamefont{Wiesendanger}},
  \bibinfo{author}{\bibfnamefont{G.}~\bibnamefont{Bihlmayer}},
  \bibnamefont{and}
  \bibinfo{author}{\bibfnamefont{S.}~\bibnamefont{Bl{\"u}gel}},
  \bibinfo{journal}{Nat. Phys.} \textbf{\bibinfo{volume}{7}},
  \bibinfo{pages}{713} (\bibinfo{year}{2011}).

\bibitem[{\citenamefont{Adams et~al.}(2012)\citenamefont{Adams, Chacon, Wagner,
  Bauer, Brandl, Pedersen, Berger, Lemmens, and Pfleiderer}}]{Adams2012}
\bibinfo{author}{\bibfnamefont{T.}~\bibnamefont{Adams}},
  \bibinfo{author}{\bibfnamefont{A.}~\bibnamefont{Chacon}},
  \bibinfo{author}{\bibfnamefont{M.}~\bibnamefont{Wagner}},
  \bibinfo{author}{\bibfnamefont{A.}~\bibnamefont{Bauer}},
  \bibinfo{author}{\bibfnamefont{G.}~\bibnamefont{Brandl}},
  \bibinfo{author}{\bibfnamefont{B.}~\bibnamefont{Pedersen}},
  \bibinfo{author}{\bibfnamefont{H.}~\bibnamefont{Berger}},
  \bibinfo{author}{\bibfnamefont{P.}~\bibnamefont{Lemmens}}, \bibnamefont{and}
  \bibinfo{author}{\bibfnamefont{C.}~\bibnamefont{Pfleiderer}},
  \bibinfo{journal}{Phys. Rev. Lett.} \textbf{\bibinfo{volume}{108}},
  \bibinfo{pages}{237204} (\bibinfo{year}{2012}).

\bibitem[{\citenamefont{Kurumaji et~al.}(2019)\citenamefont{Kurumaji, Nakajima,
  Hirschberger, Kikkawa, Yamasaki, Sagayama, Nakao, Taguchi, Arima, and
  Tokura}}]{kurumaji2019skyrmion}
\bibinfo{author}{\bibfnamefont{T.}~\bibnamefont{Kurumaji}},
  \bibinfo{author}{\bibfnamefont{T.}~\bibnamefont{Nakajima}},
  \bibinfo{author}{\bibfnamefont{M.}~\bibnamefont{Hirschberger}},
  \bibinfo{author}{\bibfnamefont{A.}~\bibnamefont{Kikkawa}},
  \bibinfo{author}{\bibfnamefont{Y.}~\bibnamefont{Yamasaki}},
  \bibinfo{author}{\bibfnamefont{H.}~\bibnamefont{Sagayama}},
  \bibinfo{author}{\bibfnamefont{H.}~\bibnamefont{Nakao}},
  \bibinfo{author}{\bibfnamefont{Y.}~\bibnamefont{Taguchi}},
  \bibinfo{author}{\bibfnamefont{T.-h.} \bibnamefont{Arima}}, \bibnamefont{and}
  \bibinfo{author}{\bibfnamefont{Y.}~\bibnamefont{Tokura}},
  \bibinfo{journal}{Science} \textbf{\bibinfo{volume}{365}},
  \bibinfo{pages}{914} (\bibinfo{year}{2019}).

\bibitem[{\citenamefont{Khanh et~al.}(2020)\citenamefont{Khanh, Nakajima, Yu,
  Gao, Shibata, Hirschberger, Yamasaki, Sagayama, Nakao, Peng
  et~al.}}]{khanh2020nanometric}
\bibinfo{author}{\bibfnamefont{N.~D.} \bibnamefont{Khanh}},
  \bibinfo{author}{\bibfnamefont{T.}~\bibnamefont{Nakajima}},
  \bibinfo{author}{\bibfnamefont{X.}~\bibnamefont{Yu}},
  \bibinfo{author}{\bibfnamefont{S.}~\bibnamefont{Gao}},
  \bibinfo{author}{\bibfnamefont{K.}~\bibnamefont{Shibata}},
  \bibinfo{author}{\bibfnamefont{M.}~\bibnamefont{Hirschberger}},
  \bibinfo{author}{\bibfnamefont{Y.}~\bibnamefont{Yamasaki}},
  \bibinfo{author}{\bibfnamefont{H.}~\bibnamefont{Sagayama}},
  \bibinfo{author}{\bibfnamefont{H.}~\bibnamefont{Nakao}},
  \bibinfo{author}{\bibfnamefont{L.}~\bibnamefont{Peng}}, \bibnamefont{et~al.},
  \bibinfo{journal}{Nat. Nanotechnol.} \textbf{\bibinfo{volume}{15}},
  \bibinfo{pages}{444} (\bibinfo{year}{2020}).

\bibitem[{\citenamefont{Tokura and
  Kanazawa}(2021)}]{Tokura_doi:10.1021/acs.chemrev.0c00297}
\bibinfo{author}{\bibfnamefont{Y.}~\bibnamefont{Tokura}} \bibnamefont{and}
  \bibinfo{author}{\bibfnamefont{N.}~\bibnamefont{Kanazawa}},
  \bibinfo{journal}{Chem. Rev.} \textbf{\bibinfo{volume}{121}},
  \bibinfo{pages}{2857} (\bibinfo{year}{2021}).

\bibitem[{\citenamefont{Dzyaloshinsky}(1958)}]{dzyaloshinsky1958thermodynamic}
\bibinfo{author}{\bibfnamefont{I.}~\bibnamefont{Dzyaloshinsky}},
  \bibinfo{journal}{J. Phys. Chem. Solids} \textbf{\bibinfo{volume}{4}},
  \bibinfo{pages}{241} (\bibinfo{year}{1958}).

\bibitem[{\citenamefont{Moriya}(1960)}]{moriya1960anisotropic}
\bibinfo{author}{\bibfnamefont{T.}~\bibnamefont{Moriya}},
  \bibinfo{journal}{Phys. Rev.} \textbf{\bibinfo{volume}{120}},
  \bibinfo{pages}{91} (\bibinfo{year}{1960}).

\bibitem[{\citenamefont{R{\"o}{\ss}ler
  et~al.}(2006)\citenamefont{R{\"o}{\ss}ler, Bogdanov, and
  Pfleiderer}}]{rossler2006spontaneous}
\bibinfo{author}{\bibfnamefont{U.~K.} \bibnamefont{R{\"o}{\ss}ler}},
  \bibinfo{author}{\bibfnamefont{A.~N.} \bibnamefont{Bogdanov}},
  \bibnamefont{and}
  \bibinfo{author}{\bibfnamefont{C.}~\bibnamefont{Pfleiderer}},
  \bibinfo{journal}{Nature} \textbf{\bibinfo{volume}{442}},
  \bibinfo{pages}{797} (\bibinfo{year}{2006}).

\bibitem[{\citenamefont{Yi et~al.}(2009)\citenamefont{Yi, Onoda, Nagaosa, and
  Han}}]{Yi_PhysRevB.80.054416}
\bibinfo{author}{\bibfnamefont{S.~D.} \bibnamefont{Yi}},
  \bibinfo{author}{\bibfnamefont{S.}~\bibnamefont{Onoda}},
  \bibinfo{author}{\bibfnamefont{N.}~\bibnamefont{Nagaosa}}, \bibnamefont{and}
  \bibinfo{author}{\bibfnamefont{J.~H.} \bibnamefont{Han}},
  \bibinfo{journal}{Phys. Rev. B} \textbf{\bibinfo{volume}{80}},
  \bibinfo{pages}{054416} (\bibinfo{year}{2009}).

\bibitem[{\citenamefont{Okubo et~al.}(2012)\citenamefont{Okubo, Chung, and
  Kawamura}}]{Okubo_PhysRevLett.108.017206}
\bibinfo{author}{\bibfnamefont{T.}~\bibnamefont{Okubo}},
  \bibinfo{author}{\bibfnamefont{S.}~\bibnamefont{Chung}}, \bibnamefont{and}
  \bibinfo{author}{\bibfnamefont{H.}~\bibnamefont{Kawamura}},
  \bibinfo{journal}{Phys. Rev. Lett.} \textbf{\bibinfo{volume}{108}},
  \bibinfo{pages}{017206} (\bibinfo{year}{2012}).

\bibitem[{\citenamefont{Leonov and Mostovoy}(2015)}]{leonov2015multiply}
\bibinfo{author}{\bibfnamefont{A.~O.} \bibnamefont{Leonov}} \bibnamefont{and}
  \bibinfo{author}{\bibfnamefont{M.}~\bibnamefont{Mostovoy}},
  \bibinfo{journal}{Nat. Commun.} \textbf{\bibinfo{volume}{6}},
  \bibinfo{pages}{8275} (\bibinfo{year}{2015}).

\bibitem[{\citenamefont{Batista et~al.}(2016)\citenamefont{Batista, Lin,
  Hayami, and Kamiya}}]{batista2016frustration}
\bibinfo{author}{\bibfnamefont{C.~D.} \bibnamefont{Batista}},
  \bibinfo{author}{\bibfnamefont{S.-Z.} \bibnamefont{Lin}},
  \bibinfo{author}{\bibfnamefont{S.}~\bibnamefont{Hayami}}, \bibnamefont{and}
  \bibinfo{author}{\bibfnamefont{Y.}~\bibnamefont{Kamiya}},
  \bibinfo{journal}{Rep. Prog. Phys.} \textbf{\bibinfo{volume}{79}},
  \bibinfo{pages}{084504} (\bibinfo{year}{2016}).

\bibitem[{\citenamefont{Hayami and
  Motome}(2021{\natexlab{a}})}]{hayami2021topological}
\bibinfo{author}{\bibfnamefont{S.}~\bibnamefont{Hayami}} \bibnamefont{and}
  \bibinfo{author}{\bibfnamefont{Y.}~\bibnamefont{Motome}},
  \bibinfo{journal}{J. Phys.: Condens. Matter} \textbf{\bibinfo{volume}{33}},
  \bibinfo{pages}{443001} (\bibinfo{year}{2021}{\natexlab{a}}).

\bibitem[{\citenamefont{Yambe and Hayami}(2022)}]{Yambe_PhysRevB.106.174437}
\bibinfo{author}{\bibfnamefont{R.}~\bibnamefont{Yambe}} \bibnamefont{and}
  \bibinfo{author}{\bibfnamefont{S.}~\bibnamefont{Hayami}},
  \bibinfo{journal}{Phys. Rev. B} \textbf{\bibinfo{volume}{106}},
  \bibinfo{pages}{174437} (\bibinfo{year}{2022}).

\bibitem[{\citenamefont{Hayami and
  Yambe}(2021{\natexlab{a}})}]{hayami2021field}
\bibinfo{author}{\bibfnamefont{S.}~\bibnamefont{Hayami}} \bibnamefont{and}
  \bibinfo{author}{\bibfnamefont{R.}~\bibnamefont{Yambe}}, \bibinfo{journal}{J.
  Phys. Soc. Jpn.} \textbf{\bibinfo{volume}{90}}, \bibinfo{pages}{073705}
  (\bibinfo{year}{2021}{\natexlab{a}}).

\bibitem[{\citenamefont{G\"obel et~al.}(2019)\citenamefont{G\"obel, Mook, Henk,
  Mertig, and Tretiakov}}]{Borge_PhysRevB.99.060407}
\bibinfo{author}{\bibfnamefont{B.}~\bibnamefont{G\"obel}},
  \bibinfo{author}{\bibfnamefont{A.}~\bibnamefont{Mook}},
  \bibinfo{author}{\bibfnamefont{J.}~\bibnamefont{Henk}},
  \bibinfo{author}{\bibfnamefont{I.}~\bibnamefont{Mertig}}, \bibnamefont{and}
  \bibinfo{author}{\bibfnamefont{O.~A.} \bibnamefont{Tretiakov}},
  \bibinfo{journal}{Phys. Rev. B} \textbf{\bibinfo{volume}{99}},
  \bibinfo{pages}{060407(R)} (\bibinfo{year}{2019}).

\bibitem[{\citenamefont{Moon et~al.}(2019)\citenamefont{Moon, Yoon, Kim, and
  Hwang}}]{Moon_PhysRevApplied.12.064054}
\bibinfo{author}{\bibfnamefont{K.-W.} \bibnamefont{Moon}},
  \bibinfo{author}{\bibfnamefont{J.}~\bibnamefont{Yoon}},
  \bibinfo{author}{\bibfnamefont{C.}~\bibnamefont{Kim}}, \bibnamefont{and}
  \bibinfo{author}{\bibfnamefont{C.}~\bibnamefont{Hwang}},
  \bibinfo{journal}{Phys. Rev. Appl.} \textbf{\bibinfo{volume}{12}},
  \bibinfo{pages}{064054} (\bibinfo{year}{2019}).

\bibitem[{\citenamefont{Hayami}(2021)}]{Hayami_PhysRevB.103.224418}
\bibinfo{author}{\bibfnamefont{S.}~\bibnamefont{Hayami}},
  \bibinfo{journal}{Phys. Rev. B} \textbf{\bibinfo{volume}{103}},
  \bibinfo{pages}{224418} (\bibinfo{year}{2021}).

\bibitem[{\citenamefont{Hayami and
  Motome}(2021{\natexlab{b}})}]{Hayami_PhysRevB.103.054422}
\bibinfo{author}{\bibfnamefont{S.}~\bibnamefont{Hayami}} \bibnamefont{and}
  \bibinfo{author}{\bibfnamefont{Y.}~\bibnamefont{Motome}},
  \bibinfo{journal}{Phys. Rev. B} \textbf{\bibinfo{volume}{103}},
  \bibinfo{pages}{054422} (\bibinfo{year}{2021}{\natexlab{b}}).

\bibitem[{\citenamefont{Hayami and
  Motome}(2021{\natexlab{c}})}]{Hayami_PhysRevB.103.024439}
\bibinfo{author}{\bibfnamefont{S.}~\bibnamefont{Hayami}} \bibnamefont{and}
  \bibinfo{author}{\bibfnamefont{Y.}~\bibnamefont{Motome}},
  \bibinfo{journal}{Phys. Rev. B} \textbf{\bibinfo{volume}{103}},
  \bibinfo{pages}{024439} (\bibinfo{year}{2021}{\natexlab{c}}).

\bibitem[{\citenamefont{Utesov}(2021)}]{Utesov_PhysRevB.103.064414}
\bibinfo{author}{\bibfnamefont{O.~I.} \bibnamefont{Utesov}},
  \bibinfo{journal}{Phys. Rev. B} \textbf{\bibinfo{volume}{103}},
  \bibinfo{pages}{064414} (\bibinfo{year}{2021}).

\bibitem[{\citenamefont{Wang et~al.}(2021)\citenamefont{Wang, Su, Lin, and
  Batista}}]{Wang_PhysRevB.103.104408}
\bibinfo{author}{\bibfnamefont{Z.}~\bibnamefont{Wang}},
  \bibinfo{author}{\bibfnamefont{Y.}~\bibnamefont{Su}},
  \bibinfo{author}{\bibfnamefont{S.-Z.} \bibnamefont{Lin}}, \bibnamefont{and}
  \bibinfo{author}{\bibfnamefont{C.~D.} \bibnamefont{Batista}},
  \bibinfo{journal}{Phys. Rev. B} \textbf{\bibinfo{volume}{103}},
  \bibinfo{pages}{104408} (\bibinfo{year}{2021}).

\bibitem[{\citenamefont{Hayami}(2022{\natexlab{a}})}]{Hayami_PhysRevB.105.174437}
\bibinfo{author}{\bibfnamefont{S.}~\bibnamefont{Hayami}},
  \bibinfo{journal}{Phys. Rev. B} \textbf{\bibinfo{volume}{105}},
  \bibinfo{pages}{174437} (\bibinfo{year}{2022}{\natexlab{a}}).

\bibitem[{\citenamefont{Yasui et~al.}(2020)\citenamefont{Yasui, Butler, Khanh,
  Hayami, Nomoto, Hanaguri, Motome, Arita, h.~Arima, Tokura
  et~al.}}]{Yasui2020imaging}
\bibinfo{author}{\bibfnamefont{Y.}~\bibnamefont{Yasui}},
  \bibinfo{author}{\bibfnamefont{C.~J.} \bibnamefont{Butler}},
  \bibinfo{author}{\bibfnamefont{N.~D.} \bibnamefont{Khanh}},
  \bibinfo{author}{\bibfnamefont{S.}~\bibnamefont{Hayami}},
  \bibinfo{author}{\bibfnamefont{T.}~\bibnamefont{Nomoto}},
  \bibinfo{author}{\bibfnamefont{T.}~\bibnamefont{Hanaguri}},
  \bibinfo{author}{\bibfnamefont{Y.}~\bibnamefont{Motome}},
  \bibinfo{author}{\bibfnamefont{R.}~\bibnamefont{Arita}},
  \bibinfo{author}{\bibfnamefont{T.}~\bibnamefont{h.~Arima}},
  \bibinfo{author}{\bibfnamefont{Y.}~\bibnamefont{Tokura}},
  \bibnamefont{et~al.}, \bibinfo{journal}{Nat. Commun.}
  \textbf{\bibinfo{volume}{11}}, \bibinfo{pages}{5925} (\bibinfo{year}{2020}).

\bibitem[{\citenamefont{Khanh et~al.}(2022)\citenamefont{Khanh, Nakajima,
  Hayami, Gao, Yamasaki, Sagayama, Nakao, Takagi, Motome, Tokura
  et~al.}}]{khanh2022zoology}
\bibinfo{author}{\bibfnamefont{N.~D.} \bibnamefont{Khanh}},
  \bibinfo{author}{\bibfnamefont{T.}~\bibnamefont{Nakajima}},
  \bibinfo{author}{\bibfnamefont{S.}~\bibnamefont{Hayami}},
  \bibinfo{author}{\bibfnamefont{S.}~\bibnamefont{Gao}},
  \bibinfo{author}{\bibfnamefont{Y.}~\bibnamefont{Yamasaki}},
  \bibinfo{author}{\bibfnamefont{H.}~\bibnamefont{Sagayama}},
  \bibinfo{author}{\bibfnamefont{H.}~\bibnamefont{Nakao}},
  \bibinfo{author}{\bibfnamefont{R.}~\bibnamefont{Takagi}},
  \bibinfo{author}{\bibfnamefont{Y.}~\bibnamefont{Motome}},
  \bibinfo{author}{\bibfnamefont{Y.}~\bibnamefont{Tokura}},
  \bibnamefont{et~al.}, \bibinfo{journal}{Adv. Sci.}
  \textbf{\bibinfo{volume}{9}}, \bibinfo{pages}{2105452}
  (\bibinfo{year}{2022}).

\bibitem[{\citenamefont{Matsuyama et~al.}(2023)\citenamefont{Matsuyama, Nomura,
  Imajo, Nomoto, Arita, Sudo, Kimata, Khanh, Takagi, Tokura
  et~al.}}]{Matsuyama_PhysRevB.107.104421}
\bibinfo{author}{\bibfnamefont{N.}~\bibnamefont{Matsuyama}},
  \bibinfo{author}{\bibfnamefont{T.}~\bibnamefont{Nomura}},
  \bibinfo{author}{\bibfnamefont{S.}~\bibnamefont{Imajo}},
  \bibinfo{author}{\bibfnamefont{T.}~\bibnamefont{Nomoto}},
  \bibinfo{author}{\bibfnamefont{R.}~\bibnamefont{Arita}},
  \bibinfo{author}{\bibfnamefont{K.}~\bibnamefont{Sudo}},
  \bibinfo{author}{\bibfnamefont{M.}~\bibnamefont{Kimata}},
  \bibinfo{author}{\bibfnamefont{N.~D.} \bibnamefont{Khanh}},
  \bibinfo{author}{\bibfnamefont{R.}~\bibnamefont{Takagi}},
  \bibinfo{author}{\bibfnamefont{Y.}~\bibnamefont{Tokura}},
  \bibnamefont{et~al.}, \bibinfo{journal}{Phys. Rev. B}
  \textbf{\bibinfo{volume}{107}}, \bibinfo{pages}{104421}
  (\bibinfo{year}{2023}).

\bibitem[{\citenamefont{Hayami and Kato}(2023)}]{hayami2023widely}
\bibinfo{author}{\bibfnamefont{S.}~\bibnamefont{Hayami}} \bibnamefont{and}
  \bibinfo{author}{\bibfnamefont{Y.}~\bibnamefont{Kato}}, \bibinfo{journal}{J.
  Magn. Magn. Mater.} \textbf{\bibinfo{volume}{571}}, \bibinfo{pages}{170547}
  (\bibinfo{year}{2023}).

\bibitem[{\citenamefont{Wood et~al.}(2023)\citenamefont{Wood, Khalyavin, Mayoh,
  Bouaziz, Hall, Holt, Orlandi, Manuel, Bl\"ugel, Staunton
  et~al.}}]{Wood_PhysRevB.107.L180402}
\bibinfo{author}{\bibfnamefont{G.~D.~A.} \bibnamefont{Wood}},
  \bibinfo{author}{\bibfnamefont{D.~D.} \bibnamefont{Khalyavin}},
  \bibinfo{author}{\bibfnamefont{D.~A.} \bibnamefont{Mayoh}},
  \bibinfo{author}{\bibfnamefont{J.}~\bibnamefont{Bouaziz}},
  \bibinfo{author}{\bibfnamefont{A.~E.} \bibnamefont{Hall}},
  \bibinfo{author}{\bibfnamefont{S.~J.~R.} \bibnamefont{Holt}},
  \bibinfo{author}{\bibfnamefont{F.}~\bibnamefont{Orlandi}},
  \bibinfo{author}{\bibfnamefont{P.}~\bibnamefont{Manuel}},
  \bibinfo{author}{\bibfnamefont{S.}~\bibnamefont{Bl\"ugel}},
  \bibinfo{author}{\bibfnamefont{J.~B.} \bibnamefont{Staunton}},
  \bibnamefont{et~al.}, \bibinfo{journal}{Phys. Rev. B}
  \textbf{\bibinfo{volume}{107}}, \bibinfo{pages}{L180402}
  (\bibinfo{year}{2023}).

\bibitem[{\citenamefont{Takagi et~al.}(2022)\citenamefont{Takagi, Matsuyama,
  Ukleev, Yu, White, Francoual, Mardegan, Hayami, Saito, Kaneko
  et~al.}}]{takagi2022square}
\bibinfo{author}{\bibfnamefont{R.}~\bibnamefont{Takagi}},
  \bibinfo{author}{\bibfnamefont{N.}~\bibnamefont{Matsuyama}},
  \bibinfo{author}{\bibfnamefont{V.}~\bibnamefont{Ukleev}},
  \bibinfo{author}{\bibfnamefont{L.}~\bibnamefont{Yu}},
  \bibinfo{author}{\bibfnamefont{J.~S.} \bibnamefont{White}},
  \bibinfo{author}{\bibfnamefont{S.}~\bibnamefont{Francoual}},
  \bibinfo{author}{\bibfnamefont{J.~R.~L.} \bibnamefont{Mardegan}},
  \bibinfo{author}{\bibfnamefont{S.}~\bibnamefont{Hayami}},
  \bibinfo{author}{\bibfnamefont{H.}~\bibnamefont{Saito}},
  \bibinfo{author}{\bibfnamefont{K.}~\bibnamefont{Kaneko}},
  \bibnamefont{et~al.}, \bibinfo{journal}{Nat. Commun.}
  \textbf{\bibinfo{volume}{13}}, \bibinfo{pages}{1472} (\bibinfo{year}{2022}).

\bibitem[{\citenamefont{Zhu et~al.}(2022)\citenamefont{Zhu, Zhang, Gawryluk,
  Zhen, Yu, Ju, Xie, Jiang, Cheng, Xu et~al.}}]{Zhu_PhysRevB.105.014423}
\bibinfo{author}{\bibfnamefont{X.~Y.} \bibnamefont{Zhu}},
  \bibinfo{author}{\bibfnamefont{H.}~\bibnamefont{Zhang}},
  \bibinfo{author}{\bibfnamefont{D.~J.} \bibnamefont{Gawryluk}},
  \bibinfo{author}{\bibfnamefont{Z.~X.} \bibnamefont{Zhen}},
  \bibinfo{author}{\bibfnamefont{B.~C.} \bibnamefont{Yu}},
  \bibinfo{author}{\bibfnamefont{S.~L.} \bibnamefont{Ju}},
  \bibinfo{author}{\bibfnamefont{W.}~\bibnamefont{Xie}},
  \bibinfo{author}{\bibfnamefont{D.~M.} \bibnamefont{Jiang}},
  \bibinfo{author}{\bibfnamefont{W.~J.} \bibnamefont{Cheng}},
  \bibinfo{author}{\bibfnamefont{Y.}~\bibnamefont{Xu}}, \bibnamefont{et~al.},
  \bibinfo{journal}{Phys. Rev. B} \textbf{\bibinfo{volume}{105}},
  \bibinfo{pages}{014423} (\bibinfo{year}{2022}).

\bibitem[{\citenamefont{Hayami}(2023)}]{hayami2023orthorhombic}
\bibinfo{author}{\bibfnamefont{S.}~\bibnamefont{Hayami}}, \bibinfo{journal}{J.
  Phys.: Mater.} \textbf{\bibinfo{volume}{6}}, \bibinfo{pages}{014006}
  (\bibinfo{year}{2023}).

\bibitem[{\citenamefont{Gen et~al.}(2023)\citenamefont{Gen, Takagi, Watanabe,
  Kitou, Sagayama, Matsuyama, Kohama, Ikeda, \ifmmode~\bar{O}\else
  \={O}\fi{}nuki, Kurumaji et~al.}}]{Gen_PhysRevB.107.L020410}
\bibinfo{author}{\bibfnamefont{M.}~\bibnamefont{Gen}},
  \bibinfo{author}{\bibfnamefont{R.}~\bibnamefont{Takagi}},
  \bibinfo{author}{\bibfnamefont{Y.}~\bibnamefont{Watanabe}},
  \bibinfo{author}{\bibfnamefont{S.}~\bibnamefont{Kitou}},
  \bibinfo{author}{\bibfnamefont{H.}~\bibnamefont{Sagayama}},
  \bibinfo{author}{\bibfnamefont{N.}~\bibnamefont{Matsuyama}},
  \bibinfo{author}{\bibfnamefont{Y.}~\bibnamefont{Kohama}},
  \bibinfo{author}{\bibfnamefont{A.}~\bibnamefont{Ikeda}},
  \bibinfo{author}{\bibfnamefont{Y.}~\bibnamefont{\ifmmode~\bar{O}\else
  \={O}\fi{}nuki}}, \bibinfo{author}{\bibfnamefont{T.}~\bibnamefont{Kurumaji}},
  \bibnamefont{et~al.}, \bibinfo{journal}{Phys. Rev. B}
  \textbf{\bibinfo{volume}{107}}, \bibinfo{pages}{L020410}
  (\bibinfo{year}{2023}).

\bibitem[{\citenamefont{Akagi et~al.}(2012)\citenamefont{Akagi, Udagawa, and
  Motome}}]{Akagi_PhysRevLett.108.096401}
\bibinfo{author}{\bibfnamefont{Y.}~\bibnamefont{Akagi}},
  \bibinfo{author}{\bibfnamefont{M.}~\bibnamefont{Udagawa}}, \bibnamefont{and}
  \bibinfo{author}{\bibfnamefont{Y.}~\bibnamefont{Motome}},
  \bibinfo{journal}{Phys. Rev. Lett.} \textbf{\bibinfo{volume}{108}},
  \bibinfo{pages}{096401} (\bibinfo{year}{2012}).

\bibitem[{\citenamefont{Hayami and Motome}(2014)}]{Hayami_PhysRevB.90.060402}
\bibinfo{author}{\bibfnamefont{S.}~\bibnamefont{Hayami}} \bibnamefont{and}
  \bibinfo{author}{\bibfnamefont{Y.}~\bibnamefont{Motome}},
  \bibinfo{journal}{Phys. Rev. B} \textbf{\bibinfo{volume}{90}},
  \bibinfo{pages}{060402(R)} (\bibinfo{year}{2014}).

\bibitem[{\citenamefont{Hayami et~al.}(2017)\citenamefont{Hayami, Ozawa, and
  Motome}}]{Hayami_PhysRevB.95.224424}
\bibinfo{author}{\bibfnamefont{S.}~\bibnamefont{Hayami}},
  \bibinfo{author}{\bibfnamefont{R.}~\bibnamefont{Ozawa}}, \bibnamefont{and}
  \bibinfo{author}{\bibfnamefont{Y.}~\bibnamefont{Motome}},
  \bibinfo{journal}{Phys. Rev. B} \textbf{\bibinfo{volume}{95}},
  \bibinfo{pages}{224424} (\bibinfo{year}{2017}).

\bibitem[{\citenamefont{Hayami and
  Motome}(2018)}]{Hayami_PhysRevLett.121.137202}
\bibinfo{author}{\bibfnamefont{S.}~\bibnamefont{Hayami}} \bibnamefont{and}
  \bibinfo{author}{\bibfnamefont{Y.}~\bibnamefont{Motome}},
  \bibinfo{journal}{Phys. Rev. Lett.} \textbf{\bibinfo{volume}{121}},
  \bibinfo{pages}{137202} (\bibinfo{year}{2018}).

\bibitem[{\citenamefont{Hayami}(2022{\natexlab{b}})}]{hayami2022multiple}
\bibinfo{author}{\bibfnamefont{S.}~\bibnamefont{Hayami}}, \bibinfo{journal}{J.
  Phys. Soc. Jpn.} \textbf{\bibinfo{volume}{91}}, \bibinfo{pages}{023705}
  (\bibinfo{year}{2022}{\natexlab{b}}).

\bibitem[{\citenamefont{Solenov et~al.}(2012)\citenamefont{Solenov, Mozyrsky,
  and Martin}}]{Solenov_PhysRevLett.108.096403}
\bibinfo{author}{\bibfnamefont{D.}~\bibnamefont{Solenov}},
  \bibinfo{author}{\bibfnamefont{D.}~\bibnamefont{Mozyrsky}}, \bibnamefont{and}
  \bibinfo{author}{\bibfnamefont{I.}~\bibnamefont{Martin}},
  \bibinfo{journal}{Phys. Rev. Lett.} \textbf{\bibinfo{volume}{108}},
  \bibinfo{pages}{096403} (\bibinfo{year}{2012}).

\bibitem[{\citenamefont{Hayami et~al.}(2016)\citenamefont{Hayami, Lin, Kamiya,
  and Batista}}]{Hayami_PhysRevB.94.174420}
\bibinfo{author}{\bibfnamefont{S.}~\bibnamefont{Hayami}},
  \bibinfo{author}{\bibfnamefont{S.-Z.} \bibnamefont{Lin}},
  \bibinfo{author}{\bibfnamefont{Y.}~\bibnamefont{Kamiya}}, \bibnamefont{and}
  \bibinfo{author}{\bibfnamefont{C.~D.} \bibnamefont{Batista}},
  \bibinfo{journal}{Phys. Rev. B} \textbf{\bibinfo{volume}{94}},
  \bibinfo{pages}{174420} (\bibinfo{year}{2016}).

\bibitem[{\citenamefont{Hayami and
  Yambe}(2020)}]{Hayami_doi:10.7566/JPSJ.89.103702}
\bibinfo{author}{\bibfnamefont{S.}~\bibnamefont{Hayami}} \bibnamefont{and}
  \bibinfo{author}{\bibfnamefont{R.}~\bibnamefont{Yambe}}, \bibinfo{journal}{J.
  Phys. Soc. Jpn.} \textbf{\bibinfo{volume}{89}}, \bibinfo{pages}{103702}
  (\bibinfo{year}{2020}).

\bibitem[{\citenamefont{Hayami and Yambe}(2022)}]{Hayami_PhysRevB.105.104428}
\bibinfo{author}{\bibfnamefont{S.}~\bibnamefont{Hayami}} \bibnamefont{and}
  \bibinfo{author}{\bibfnamefont{R.}~\bibnamefont{Yambe}},
  \bibinfo{journal}{Phys. Rev. B} \textbf{\bibinfo{volume}{105}},
  \bibinfo{pages}{104428} (\bibinfo{year}{2022}).

\bibitem[{\citenamefont{Hayami and
  Yambe}(2021{\natexlab{b}})}]{Hayami_PhysRevResearch.3.043158}
\bibinfo{author}{\bibfnamefont{S.}~\bibnamefont{Hayami}} \bibnamefont{and}
  \bibinfo{author}{\bibfnamefont{R.}~\bibnamefont{Yambe}},
  \bibinfo{journal}{Phys. Rev. Research} \textbf{\bibinfo{volume}{3}},
  \bibinfo{pages}{043158} (\bibinfo{year}{2021}{\natexlab{b}}).

\bibitem[{\citenamefont{Hayami}(2020)}]{hayami2020multiple}
\bibinfo{author}{\bibfnamefont{S.}~\bibnamefont{Hayami}}, \bibinfo{journal}{J.
  Magn. Magn. Mater.} \textbf{\bibinfo{volume}{513}}, \bibinfo{pages}{167181}
  (\bibinfo{year}{2020}).

\bibitem[{\citenamefont{Takagi et~al.}(2018)\citenamefont{Takagi, White,
  Hayami, Arita, Honecker, R{\o}nnow, Tokura, and Seki}}]{takagi2018multiple}
\bibinfo{author}{\bibfnamefont{R.}~\bibnamefont{Takagi}},
  \bibinfo{author}{\bibfnamefont{J.}~\bibnamefont{White}},
  \bibinfo{author}{\bibfnamefont{S.}~\bibnamefont{Hayami}},
  \bibinfo{author}{\bibfnamefont{R.}~\bibnamefont{Arita}},
  \bibinfo{author}{\bibfnamefont{D.}~\bibnamefont{Honecker}},
  \bibinfo{author}{\bibfnamefont{H.}~\bibnamefont{R{\o}nnow}},
  \bibinfo{author}{\bibfnamefont{Y.}~\bibnamefont{Tokura}}, \bibnamefont{and}
  \bibinfo{author}{\bibfnamefont{S.}~\bibnamefont{Seki}},
  \bibinfo{journal}{Sci. Adv.} \textbf{\bibinfo{volume}{4}},
  \bibinfo{pages}{eaau3402} (\bibinfo{year}{2018}).

\end{thebibliography}
\end{document}